\documentclass[iop]{emulateapj}   % looks like an ApJ article

\usepackage{graphicx}
\usepackage{color}
\usepackage{amsmath}
\usepackage{hyperref, breakurl}
\usepackage{epstopdf}
\usepackage{natbib}

\newcommand{\OVI}{\ion{O}{6}}
\newcommand{\OVII}{\ion{O}{7}}
\newcommand{\OVIII}{\ion{O}{8}}
\newcommand{\NeVIII}{\ion{Ne}{8}}
\newcommand{\lya}{Ly$\alpha$}
\newcommand{\HI}{\ion{H}{1}}
\newcommand{\dndz}{$d{\cal N}/dz$}

\begin{document}

\author{Hilary Egan\altaffilmark{1,3}}
\author{Britton D. Smith\altaffilmark{1,4}}
\author{Brian W. O'Shea\altaffilmark{1,2}}
\author{J. Michael Shull\altaffilmark{3}}

\altaffiltext{1}{Department of Physics and Astronomy, Michigan State University, East Lansing, MI 48824, USA}
\altaffiltext{2}{Lyman Briggs College and Institute for Cyber-Enabled Research, Michigan State University, East Lansing, MI 48824, USA}
\altaffiltext{3}{Center for Astrophysics and Space Astronomy,
  Department of Astrophysical and Planetary Sciences, University of Colorado, Boulder, CO 80309, USA}
\altaffiltext{4}{Institute for Astronomy, University of Edinburgh, Royal Observatory,
	Edinburgh EH9 3HJ, UK}

\email{eganhila@msu.edu}

\title{Bringing Simulation and Observation Together to Better
  Understand the Intergalactic Medium}

\begin{abstract}
The methods by which one characterizes the distribution of matter in
cosmological simulations is intrinsically different from how one
performs the same task observationally.  In this paper, we make
substantial steps towards comparing simulations and observations of the
intergalactic medium (IGM) in a more sensible way.  We present a
pipeline that generates and fits synthetic QSO absorption spectra
using sight lines cast through a cosmological simulation, and
simultaneously identifies structure by directly analyzing the
variations in \HI\ and \OVI\ number density. We compare synthetic
absorption spectra with a less observationally motivated, but more
straightforward density threshold-based method for finding absorbers. Our efforts focus on
\HI\ and \OVI\ to better characterize the warm/hot intergalactic
medium, a subset of the IGM that is challenging to conclusively
identify observationally.  We find that the two methods trace roughly
the same quantities of \HI\ and \OVI\ above observable column density limits,
but the synthetic spectra typically identify more
substructure in absorbers.  We use both methods to characterize \HI\
and \OVI\ absorber properties.  We find that both integrated and
differential column
density distributions from both methods generally agree with observations. The
distribution of Doppler parameters between the two methods are similar for \lya\ and compare
reasonably with observational results, but while the two
  methods agree with each other with \OVI\ systems, they both are systematically different from
  observations.
  We find a strong
correlation between \OVI\ baryon fraction and \OVI\ column density. We
also discuss a possible bimodality in the temperature distribution of
the gas traced by \OVI.
\end{abstract}

\section{Introduction}

It is now well accepted that a large portion of all baryons at $z =
0$, up to $\sim60$\%, exist in a phase that is extremely difficult to
detect.  Unlike at high redshift ($z \gtrsim 4$) where nearly all
baryons are accountable through \lya\ absorption 
\citep{1997ApJ...489....7R, 2009and..book..419P}, only $\sim30$\% are
observed to be associated with the cool, photoionized Lyman alpha
forest today \citep{1992MNRAS.258P..14P, 1994MNRAS.267...13B,
1998ApJ...503..518F, 2004ApJ...616..643F, danforth2008, shull2012},
with only an additional $\sim10$\% in galaxies.  However, it has been
shown by cosmological simulations \citep[e.g.][]{1999ApJ...514....1C,
1999ApJ...511..521D, 2001ApJ...552..473D, britton2011} that most of these ``missing
baryons'' still exist in the IGM, but have simply been heated through
a combination of accretion shocks from structure formation and
feedback from galaxies to temperatures too hot ($\sim 10^{5-7}$ K) to
be easily traceable with neutral hydrogen.  Because of these high
temperatures, this so-called warm/hot intergalactic medium (WHIM) can
only be traced through absorption lines of highly ionized metals, such
as \OVI\ and \NeVIII\ at ultraviolet wavelengths, and \OVII\ and
\OVIII\ in the X-ray.  To date, the \OVI\ doublet at
$\lambda\lambda$1032, 1038~\AA\ has proven to be the most fruitful of
these detection methods, with numerous large samples having been
compiled \citep{2005ApJ...624..555D, danforth2008,
2006ApJ...640..716D, tripp2008, 2008ApJS..179...37T,
2008ApJ...683...22T, tilton2012, 2014arXiv1402.2655D, 2014ApJS..212....8S}, providing an excellent benchmark for
simulations.

Numerous simulation studies have succeeded at reproducing the observed
number density per unit redshift \dndz\ of \OVI\ absorbers
\citep[][]{2006ApJ...650..573C, 2009MNRAS.395.1875O,
britton2011, 2011ApJ...731...11C, teppergarcia2011,
2012MNRAS.420..829O,2012ApJ...753...17C}, despite making significantly different
predictions as to the physical conditions of the associated gas.  For
example, \citet{2009MNRAS.395.1875O} and \citet{2012MNRAS.420..829O}
find that an overwhelming majority of \OVI\ absorbers are associated
with cool, photoionized gas with a mean temperature of $\sim15,000$ K.
On the other hand, \citet{britton2011} and \citet{shull2012} find that \OVI\ absorbers come
from gas in two distinct temperature regimes -- a photoionized
component at $T \sim 30,000$ K and a hotter, collisionally ionized
component at $T \sim 300,000$ K, comprising $\sim35$\% and $\sim55$\%
of all \OVI\ absorption, respectively (with the remaining 10\%
contained in collapsed structures).  \citet{teppergarcia2011} find a
similar distribution of \OVI\ as \citet{britton2011} in their
simulated volume, but \OVI\ absorption is
biased towards detecting the hotter, collisionally ionized phase.  Some of these
disagreements may arise from differences in the simulations, such as
the methods by which metals are injected and mixed into the
intergalactic medium by the feedback scheme, but
another potential source of difference is the method in which
synthetic \OVI\ absorbers are created and analyzed in these works.
\citet{2009MNRAS.395.1875O}, \citet{teppergarcia2011}, and
\citet{2012MNRAS.420..829O} create synthetic absorption spectra which
are then analyzed in a fashion similar to actual observations.
However, \citet{britton2011} create absorbers simply by summing the
total column density along a small section of a randomly oriented line
of sight in their simulation box.

In this paper we present a re-analysis of a simulation from
\citet{britton2011}, with the intent being to better understand the
correspondence between absorption lines and the physical conditions
with which they are associated, and to better understand possible systematics
between the usage of different means of extracting absorber data
from simulations.  To accomplish this, we create a
sophisticated pipeline for fitting absorption-line spectra that can be
used for both synthetic and actual absorption spectra.  Our primary
goal is to understand the systematic differences between quantities
inferred from absorption lines in synthetic spectra and quantities 
derived more directly from the simulation outputs.
In Section~\ref{sec:sim}, we describe
briefly the simulation upon which our analysis is focused.  In
Section~\ref{sec:methods}, we detail three different methods of
varying sophistication that we use to create synthetic absorber
samples.  In Section~\ref{sec:results}, we examine a range of
properties of the synthetic \lya\ and \OVI\ absorbers created using
each of these three methods, comparing each method both to the other
two and to observations of the statistical properties of real quasar
absorption line systems.  We provide a discussion of our results in
Section~\ref{sec:discussion}, and summarize in
Section~\ref{sec:summary}.

\section{Simulation}
\label{sec:sim}

In this work, we analyze a simulation performed with the open source
cosmological adaptive mesh refinement + N-body code,
\texttt{Enzo}\footnote{\url{http://enzo-project.org}}.
\citep{2013arXiv1307.2265T, 2004astro.ph..3044O, 2007arXiv0705.1556N}.
The simulation, which is run 50\_1024\_2 from \citet{britton2011}, has
a comoving box size of 50 Mpc/$h$ with 1024$^{3}$ grid cells and dark
matter particles, corresponding to a comoving spatial resolution of 49
kpc/$h$ and a dark matter particle mass of $7\times10^{6}\ M_{\odot}$.
The simulation includes the metallicity dependent radiative cooling
method of \citet{2008MNRAS.385.1443S}, modified to include a
metagalactic UV background and a modified version of the star
formation and feedback method of \citet{1992ApJ...399L.113C} that
injects stellar feedback into a $3\times3\times3$ cube of grid cells
centered on the star particle (referred to in \citet{britton2011} as
the ``distributed'' method). Note that the distributed feedback method
does not take into account a physical scale, it suffers from the same
overproduction of stars and metals as its predecessor when used with higher 
resolution simulations. However,  \citet{britton2011} have shown that this
simulation is able to reasonably reproduce both the observed global
star formation history \citep{0004-637X-651-1-142} and the number density of \OVI\ absorbers per
unit redshift (\dndz) in the redshift range $0 \le z \le 0.4$.
\citet{britton2011} performed simulations using two different stellar
feedback models.  However, since the goal of this work is to
understand how the statistics of synthetic absorbers are affected by
the method with which they are produced, we choose to focus on just the
single simulation from \citet{britton2011} that best matched
observations.

\section{Approaches to Identifying IGM Absorbers}
    \label{sec:methods}

Using the simulations described in Section \ref{sec:sim} and the
\texttt{yt} analysis toolkit\footnote{\url{http://yt-project.org}}
\citep{yt}, we generate 2000 pencil-beam
data arrays meant to be representative of QSO sight lines. Each line
of sight extends from $z=0$ to $z=0.4$ and has a random starting
position in $(x,y,z)$ and random orientation in $(\theta,\phi)$. As
the physical distance corresponding to 
$\Delta z = 0.4$ is much larger than the extent of the simulation box
size at low redshifts, the line of sight generation tool stacks rays from multiple
epochs together. This is accomplished by pre-calculating the number of
data outputs and the corresponding position in redshift space needed
to traverse the full length of the desired sample.  All pertinent
information is recorded for each cell intersected by a line of sight,
including baryon density, metallicity, ion densities, ion fractions,
and temperature.  The pathlength $dl$ of the ray as it passes through
the cell is also recorded; this length is not constant given the
random orientation of the sight line, but is on the order of the resolution of the simulation.

\begin{figure}
\begin{center}
\includegraphics[width=.45\textwidth]{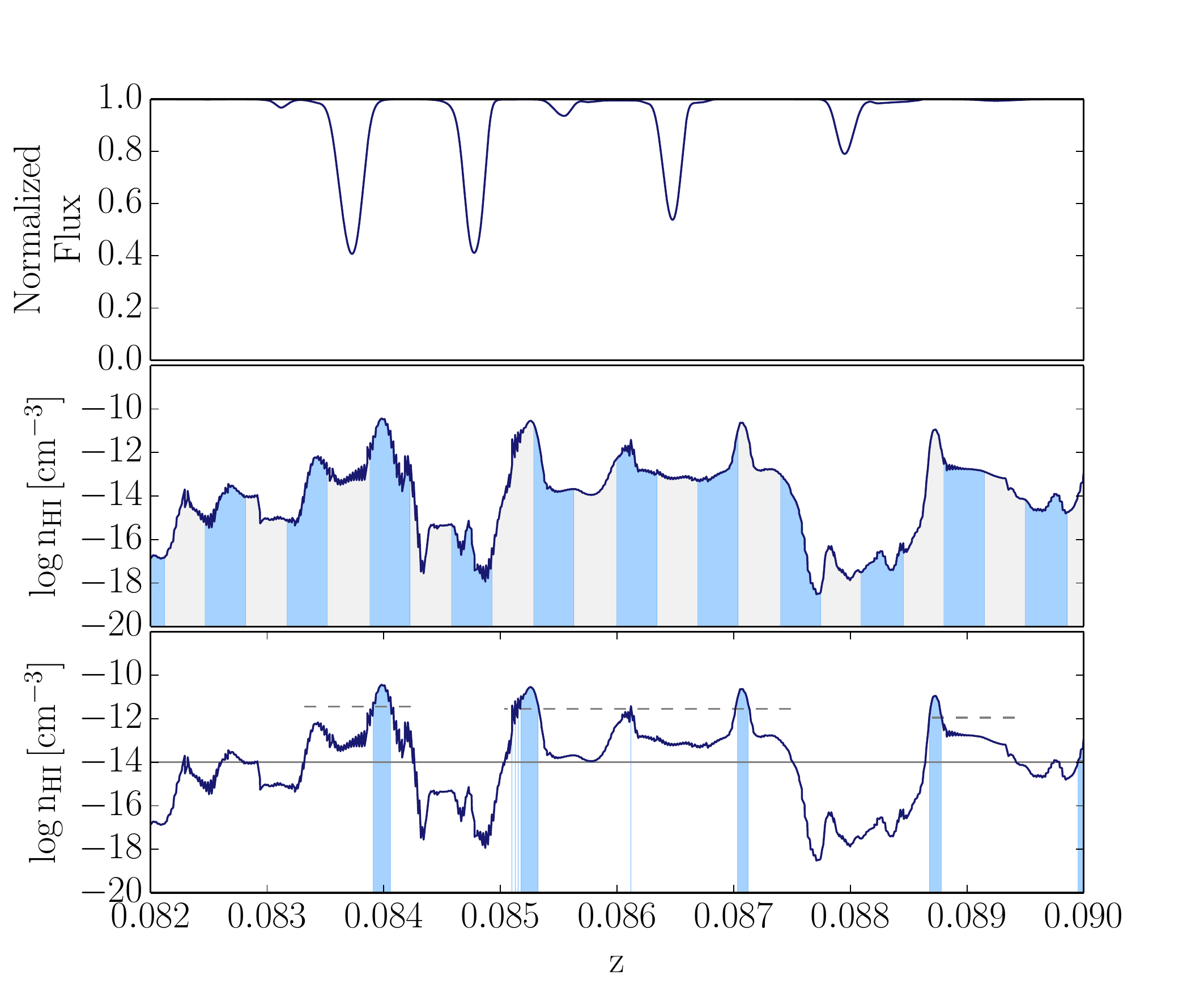}
\caption{Schematic example of three different methods for
  finding absorbers in mock QSO lines of sight. In this figure
  we examine \HI; this applies equally well to any
  species. \emph{Top row:} The ``spectral method.'' A synthetic spectrum is
  generated using the region of number density as a function of
  redshift  (with density fields shown in the bottom two rows). Normalized flux plotted against
  redshift appears shifted in redshift relative to peaks in
  number density due to line-of-sight
  velocities. \emph{Middle row:} The ``cut method.''  \HI\ number density as
  a function of redshift along the line of sight. Each shaded
  region of fixed redshift width is a separate absorber.
  \emph{Bottom row:}  The ``contour method.''  The same region of \HI\ number
  density as a function of redshift as above, but regions are identified as spatially contiguous cells above the mean number density of the species. The mean number density is given by the solid line at $\mathrm{n_{HI}}=10^{-14}$~cm$^{-3}$. The dashed line is the characteristic density for each region above the mean number density. The spatially contiguous cells identified with number densities above the characteristic densities (shaded in blue) are then identified as absorbers. }
\label{fig:methods}
\end{center}
\end{figure}

 The methods with which observers find structure in the universe is
intrinsically different than the ways simulators
approach the same problem. Even after generating lines of sight,
simulators still have access to all the information retained by the
pixels along the line of sight, or `lixels' as per \citet{britton2011},
from temperature to specific number densities. Observationally, we are
limited to interpreting these quantities through indirect methods such
as fitting absorption lines, apparent optical depth, or pixel optical
depth. Each of these approaches is susceptible to its own systematic
effects that may make it difficult to compare them. By performing both
an observationally motivated analysis technique as well as a computational
one on the same lines of sight we can better correlate
the products of simulations with actual observations. Thus, in the
following sections we will detail several methods with which we
analyzed our synthetic lines of sight; one, the `spectral' method,
that uses only absorption line fits of synthetic spectra 
and two approaches, the `contour' and
`cut' methods, which utilize cell-by-cell output. A sample of the
results of identification of absorbers using these methods can be seen
in Figure~\ref{fig:methods}. The relevant number densities ($\mathrm{n_{HI}}$,
$\mathrm{n_{OVI}}$) are computed in both photo- and collisional ionization equilibrium, with the former assuming a uniform \citet{2001cghr.confE..64H} radiation background.

For each method, after generating a full set of absorbers we filter by column density and only include \HI\ absorbers with column densities N (cm$^{-2}$) in the range $12.5\leq \log \mathrm{N_{HI}} \leq 17.2$ and \OVI\ absorbers with column densities in the range $12.5\leq \log \mathrm{N_{OVI}} \leq 15.2$. We restrict our column density range on the higher end to reflect the absence of radiative transfer in our simulation. On the low end we restrict our column density range to reflect observable limits.

\subsection{Spectral Method}

In order to mimic true observations with the data generated via
simulation, a synthetic QSO absorption spectrum was created and
then fitted for each line of sight. The fitted components then
determined the absorber -- in particular, the absorber's column density and
doppler-value. This method will henceforth be referred to as the
``spectral method.''

\subsubsection{Creating Absorption Spectra}

For each of the 2000 simulated QSO sight lines we create a corresponding
absorption spectrum by considering the effects of absorption for each 
lixel of gas. The optical depth for a cloud of gas with column density N of a
wavelength $\lambda$ is given by $\tau_i(\lambda) = C_i N a H(a,x(\lambda))$,
where $x=(\lambda-\lambda_o)/\lambda_o$, $a=\Gamma_i/(4\pi \Delta \nu_D)$, and $C_i$
is a transition-dependent constant. Here $\Delta  \nu_D = b_{th}/(c\lambda_i)$ and 
$\Gamma_i$ is the reciprocal of the mean lifetime of the transition. The Voigt-Hjerting function, $H$, is given by 

\begin{equation}
  H(a,x) = \frac{a}{\pi}\int_{-\infty}^{\infty}\frac{e^{-y^2}}{(x-y)^2+a^2}dy\;,
  \label{eq:voigt_long}
\end{equation}

where $y=v/b_{th}$. The central wavelength $\lambda_o$ is given by 

\begin{equation}
  \lambda_o = \lambda_i(1+z)(1+v_{los}/c) \;,
\end{equation}

where $\lambda_i$ is the rest wavelength of the transition for species $i$,
$z$ is the redshift of the lixel generating the profile, and $v_{los}$ is
the peculiar velocity of the lixel generating the profile. 

By defining a Doppler parameter as in Equation \eqref{eq:bth}, we
can define and $\Delta \lambda_D = (b_{th}/c)\lambda_i$ 
where $\lambda_i$ is the resonant wavelength
of the corresponding transition given by $\lambda_i = hc/E_i$. The
thermal Doppler parameter $b_{th}$ quantifies the width of a
single line:

\begin{equation}
  b_{th}=\sqrt{2 k_B T/m_i} \;.
  \label{eq:bth}
\end{equation}

The resulting normalized flux is then given by $F(\tau) =
e^{-\tau}$. The absorption feature generated in this fashion is
known as the Voigt profile. One of these profiles is generated for
each lixel using the algorithm detailed by \citet{armstrong1967}. 

The spectra were generated at a resolving power of
$R=\lambda/\Delta\lambda=10^{5}$. This is substantially higher
resolution than a typical IGM observation using Hubble's Cosmic
Origins Spectrograph (COS), which has resolving power of $R=2,000$ (G140L) and
$18,000$ (G130M); however, the simulated spectra can easily be degraded
to any desired lower resolution, and noise can be added.

As each lixel generates its own Voigt profile, large scale
features in the spectra are typically composed of contributions
from multiple lixels and their corresponding column densities.
This effect can be seen in Figure~\ref{fig:methods} where, despite
a wide range of \HI\ number densities over a number of cells, only four
significant spectral features are visible. This causes a
degeneracy in our notion of b-value, since broadening can occur as a
result of disordered line-of-sight motion within a cloud of gas (as a
result of, e.g., turbulence) in addition to
the broadening due to thermal motion. Cells with different
velocities create Voigt profiles at slightly different wavelengths
due to the Doppler effect and, as a result, the overall line to
looks broader. This broadening is typically indistinguishable from the
broadening due to thermal motion.

In generation of the spectrum as many ions as desired
can be included. Fitting a single ion at a time removes any
ambiguity of what ion is being analyzed, so all further analysis
was completed in this manner. Our spectra were also fit without
the effects of noise as an initial test of our methods, though we
note that the method works with noisy spectra as well. 

\subsubsection{Fitting Absorption Spectra}     
Due to the challenge of fitting a large number of synthetic
spectra, we developed an automated line-fitting pipeline. This
system takes the synthetic spectra and fits a series of Voigt
profiles to each spectrum in a completely automated
fashion. To automatically identify absorbers, each contiguous
region with normalized flux below an adjustable cut-off
($F/F_{cont}=0.99$) is identified. This cut-off was chosen
primarily to underestimate the observable limit due to noise
limitations, without over-fitting very shallow and hard-to-detect lines. Lowering this value by a few percent does not
significantly affect results. Raising the value begins to
substantially increase computational cost as regions become
larger and a typical region contains more lines; however, the
results are not typically affected as a region is generally
considered successfully fit regardless of the inclusion of
the smaller fluctuations.

Exceptionally large complexes ($\Delta \lambda >50\,\r{A}$) with multiple lines
are broken up at points of minimum absorption, if that minimum
point has a flux within a few percent of the adjustable
cutoff. This is done to prevent very large regions from
dramatically slowing down the fitting procedure due to the
difficulty of optimizing the fits of many lines
simultaneously.

Each region is then taken in turn from minimum to maximum
wavelength and fit by iteratively adding and adjusting Voigt
profile parameters (column density, broadening value, and
redshift) using the least squares method until the total
$\chi^2$ error is smaller than an acceptable fractional error threshold
($10^{-4}$) multiplied by the number of points in the region
in units of normalized flux. We attempt to fit up to 8
individual components simultaneously. If this is
unsuccessful or if the error begins to increase upon adding
more components, the fit is accepted if the total error
is within two orders of magnitude of the desired error. Given
the stringent conditions used for our standard fitting procedure, these fits are
typically quite good, and would certainly be acceptable if noise
were included in our synthetic spectra (which we defer to a later paper).
Changing the value of the acceptable fractional error threshold 
by an order of magnitude has little effect on the overall results.

In cases of ions that create more than one line (as in the
\OVI\ doublet), the lower wavelength line is fit to
the region as, in the case of Li-like ions (2s-2p), this is the stronger
line. We then attempt to fit the higher wavelength line
with the parameters calculated for the lower wavelength
counterpart and if the resulting total fit has a low enough
error then it is accepted. This allows some amount of leeway
for blanketed line identification where a large line occurs in
the same wavelength space as another smaller line, effectively
hiding the smaller line.

\begin{figure}
  \includegraphics[width=0.45\textwidth]{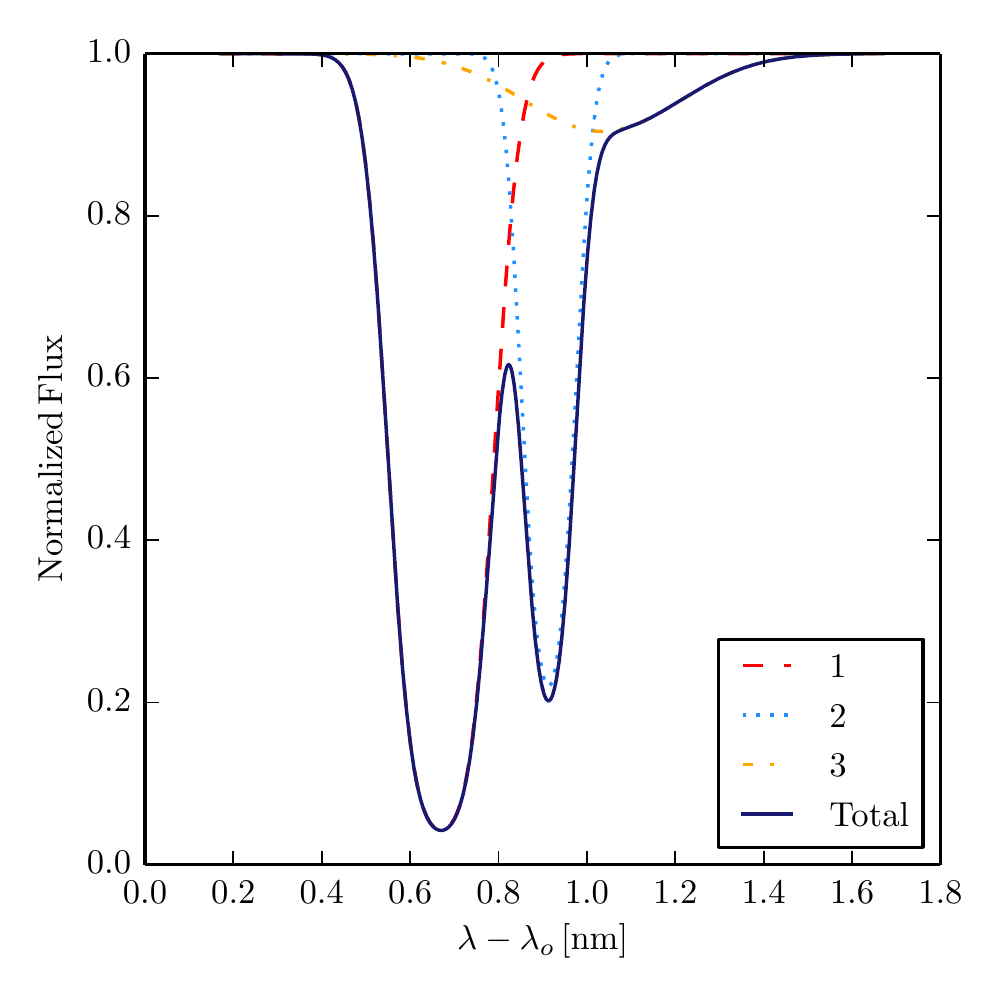}
  \caption{An example \lya\ fit for a region of wavelength space where
the best fit consists of multiple line components.  The x-axis
indicates wavelength difference measured from an arbitrary reference
wavelength, and the y-axis indicates normalized flux.  The solid blue
line indicates the total normalized flux generated from a range of
\HI\ number density along a line of sight, while the dashed lines
indicate components found for the fit.}
  \label{fig:fitex}
\end{figure}

An example of this fitting procedure is shown in Figure~\ref{fig:fitex}. 
Once the region is identified, a single line
is optimized to fit the whole region. Given the structure of
the region, a single line is insufficient to constrain the
region and the difference between the fit and the data is
still larger than allowed by the average error per point. Thus,
another line is added and the region is fit using two lines,
with three free parameters, N, $b$, and $z$. This fit is again not
good enough so this procedure is repeated now with three lines for
a total of nine free parameters. The three-line fit then satisfies the
average error per point bound and the region is considered
successfully fit.

In regions where the regular fitting tools fail to perform
properly and the line in consideration is \lya, a more
robust approach is used. Absorbers with low temperatures
($T\leq10^4$ K) and relatively high column densities
($N_{HI}\sim 10^{18}$ cm$^{-2}$) generate Voigt profiles
with damping wings. Typical parameters that
are appropriate for the large majority of absorbers often fail
to converge to a good fit for absorbers with these properties.
When such a region is identified, a separate set of widely
varied initial temperatures and column densities are tried,
which allows for a more accurate fit at the expense of
substantially increased computational cost. In
this paper we primarily examine \HI\ and \OVI\ absorption. Most
of the challenging fits are of \HI\ absorbers, due to their more
varied (and higher) column densities and environments. 

%Although the degeneracy between broadening due to thermal motion and
%random motion cannot generally be broken using a single species, we
%attempt a naive distinction when post-processing our simulated spectra
%to examine \OVI. We begin by finding the temperature below which all
%but $1\%$ of oxygen is ionized to this level or above. We then assign
%an appropriate maximum $b_{th}$ value using this temperature cutoff,
%and any $b$ values found in excess of this value are partially
%attributed to a non-thermal component such that the total $b$ remains
%constant and $b^2=b_{th}^2+b_{nth}^2$.  We do not attempt a similar
%distinction for \lya\ because, given the high concentration of hydrogen
%in the universe, a neutral fraction of $1\%$ or less can still
%generate quite substantial absorption line systems.

\subsection{Cut Method}
\label{sec:cut}
Instead of creating absorption spectra, simulators have the
opportunity to examine exactly how number density and other
physical quantities vary with redshift along a line of sight.
However, in order to compare with observation, it is necessary
to break up this continuous variation into discrete absorbing
structures. A simple approach, henceforth referred to as the
`cut method,' was employed by  \citet{britton2011}
in a previous analysis of the simulations examined in this
paper. In this method, a given line of sight is cut into
pieces of constant length in redshift space where the cells
between each cut contribute to that absorber. The sampling
resolution ($\lambda/\delta\lambda$) is set at 5000; this
allows smoothing of the region over multiple lixels while
maintaining a constant resolution in the range of
observation. A change of resolution by an order of magnitude 
in either direction does not significantly affect
the results. 

The column density for ion x is given by the sum over cells
$N_x=\sum dl_i n_{x,i}$, where $dl_i$ is the pathlength and
$n_{x,i}$ is the number density of ion x in a given absorber species in a given cell.
The b-value is comprised of a thermal and a non-thermal
component for a total b-value given by
$b=\sqrt{b_{th}^2+(\sigma_{v})^2}$.  The thermal component is
the same as the spectral thermal Doppler parameter,
$b_{th}=\sqrt{2kT/m_i}$. To approximate the non-thermal
component of broadening we find the variance in the line of
sight velocities weighted by column density of each cell. All
other relevant quantities such as ion fraction, metallicity,
and temperature are found via column density-weighted average for all cells contributing to the
absorber. 

This method is simple because it requires no additional
analysis beyond dividing the line of sight properties into
equal redshift bins; however, it does not identify
physical structures at all. Large absorbers may happen to fall
along the boundary of two redshift bins as seen at $z\sim0.085$
in Figure~\ref{fig:methods}. This may cause a single region to
be interpreted as multiple absorbers of lower column density
and distort the inferred physical conditions associated with
the absorber. Although properties dominated by the sharp peaks
in number density may still hold, this method cannot be
expected to give good intuition for the physical nature of a
structure because, quite simply, the method of defining
absorbers in not based on physics, but rather numerically
convenient quantities.

\subsection{Contour Method}
\label{sec:contour}
\begin{figure}
\begin{center}
\includegraphics[width=.45\textwidth]{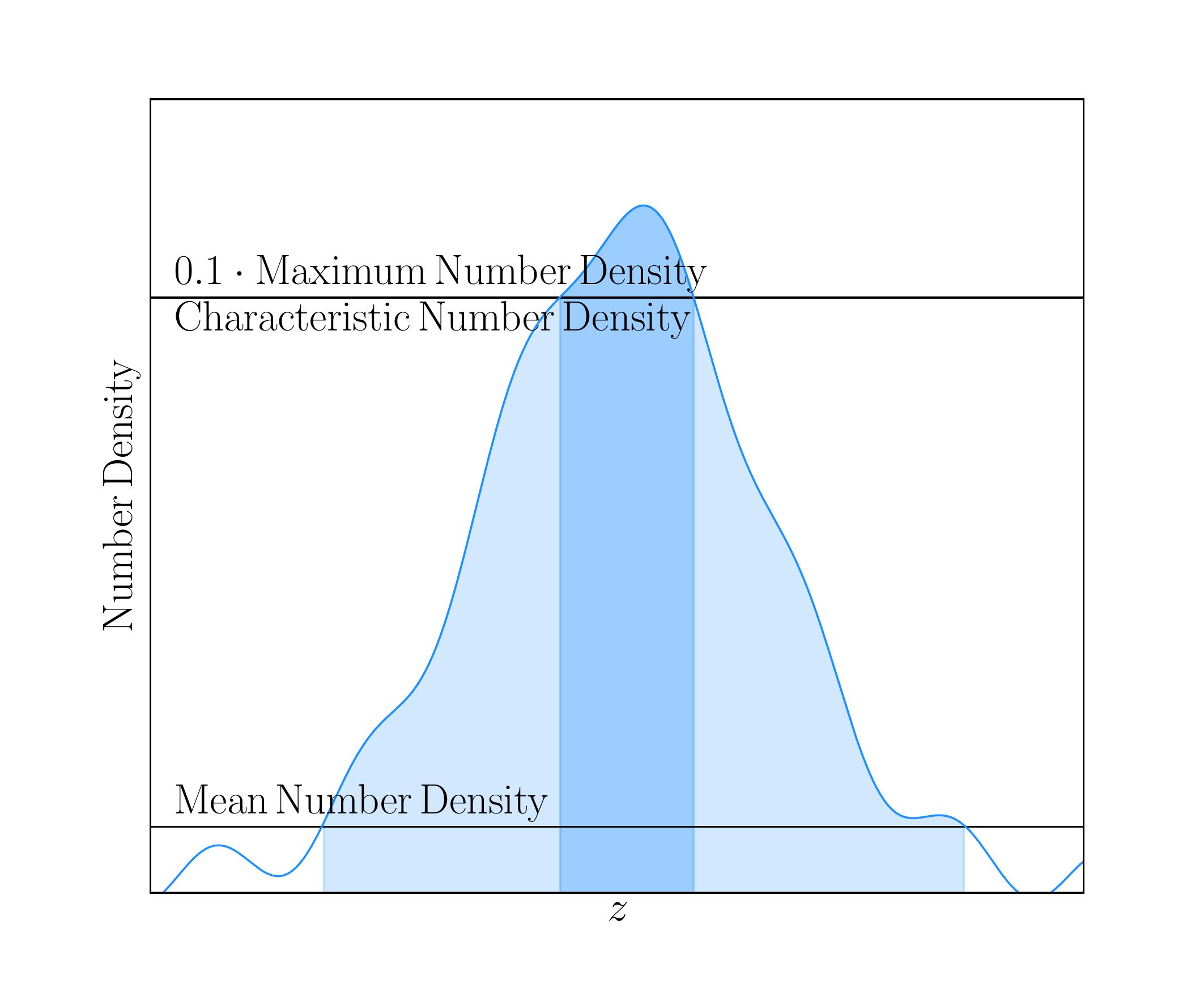}
\caption{Schematic example of the ``contour method''. Initially, the light blue region is identified as a spatially contiguous group of cells above the mean number density (\emph{Bottom Line}). Within this region the maximum number density is identified. The characteristic number density (\emph{Top line}) of this region is then calculated as some fraction of the maximum number density. The absorber for this region is then identified as the spatially contguous subset of cells with number densities greater than the characteristic number density.}
\label{fig:schematic_contour}
\end{center}
\end{figure}

The `contour method' of defining IGM absorbers was developed
in response to the limitations of the cut method described in
Section~\ref{sec:cut}. Instead of cutting blindly by redshift,
regions are identified where groups of spatially contiguous cells
all have number densities above the mean number density of a given 
species. A discussion of identifying the mean number density of each
species is given in Section~\ref{sec:cutoff}, but the values adopted
for the remained of this work are $\mathrm{n_{HI} = 10^{-14}~cm^{-3}}$ 
and $\mathrm{n_{OVI} = 10^{-22}~cm^{-3}}$.
For each region, a peak species number density is 
determined. An absorber is then quantified as spatially contiguous 
cells within the original region  with number densities 
above a characteristic density, set by some
fraction of the peak number density. For most of the following analysis
we chose the characteristic density as 0.5 times the
the peak number density, but the effects of varying this fraction are 
investigated in section~\ref{sec:cutoff}. This process is illustrated
in Figure~\ref{fig:schematic_contour}. Multiple absorbers can be identified 
in a single region as illustrated in Figure~\ref{fig:methods} at 
$z \simeq 0.086$. By setting a characteristic density for each region 
rather than a single number density cutoff for the whole species we 
ensure that only the cells that maximally contribute to a given absorbing 
feature as selected as part of the absorber, while still identifying low 
column absorbers. 

All properties of these absorbers are found using the 
same methodology as was used for the cut method. The identification 
of the regions in this way isolates large structures in a 
physically-motivated way that preserves overall structure. 

%A sample of the results of identification of absorbers using
%this method can be seen in the bottom panel of Figure
%\ref{fig:methods}. The solid line shows $n_{HI}$ along the line of
%sight and the dashed gray line at $\log_{10} n_{HI}\sim-14$ indicates the
%number density cutoff. The three shaded regions each consist of a set
%of spatially contiguous cells with number densities that all fall
%above this cutoff. Other smaller regions, such as those between the
%absorbers located at $z\sim0.044$ and $z\sim0.045$, were also
%identified but were not used for further analysis as the total column
%densities of the regions are less than $N_{HI}=10^{12.5}$
%cm$^{-2}$. This limit was chosen as it is comparable to the
%current observable limit; currently, COS cannot detect \lya\ absorbers below
%N$_{HI} \sim10^{12.5}$~cm$^{-2}$.

\section{Results}
\label{sec:results}

\subsection{Choosing an Appropriate Contour Cut-off Density}
\label{sec:cutoff}
As discussed in Section~\ref{sec:contour}, for each spatially
contiguous region of cells above the mean number density
the characteristic density is calculated as some fraction 
of the peak number density of the region. Absorbers are then 
determined by spatially contiguous cells above the characteristic
density. Thus, lowering or raising the characteristic density 
fraction could affect the number of absorbers
and the extent of individual absorbers. As the effects of a poor
choice in characteristic density fraction are not immediately obvious,
appropriate care must be taken to select a reasonable value.

\begin{figure*}
\begin{center}
    \includegraphics{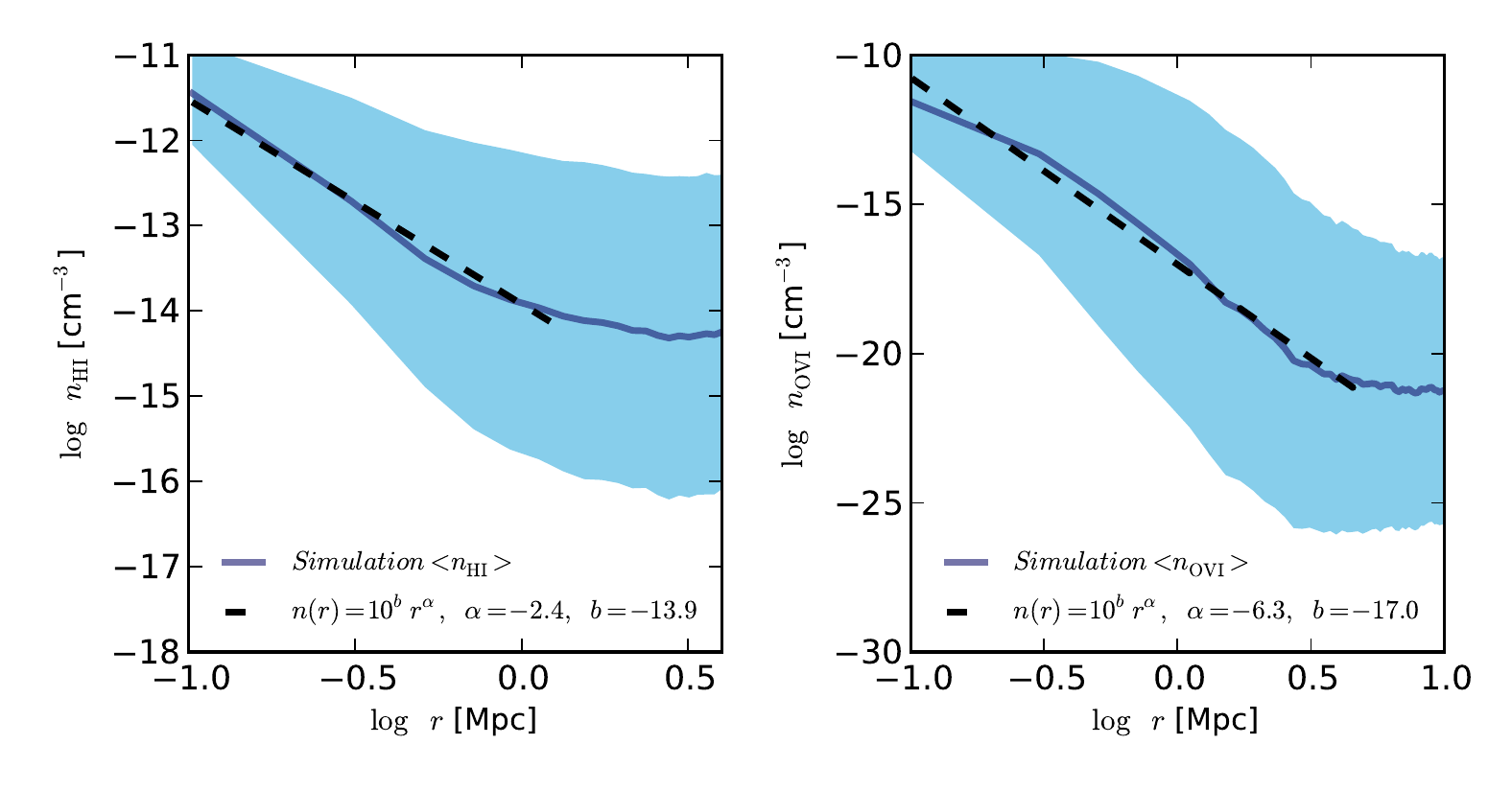}
    \end{center}
  \caption{Average number density as a function of distance along the
    line of sight from the center of an absorber.  The shaded region
    shows $1\sigma$ deviations. All absorbers with column density in
    range ($10^{12}-10^{13}$~cm$^{-2}$) for \HI\ (left) and ($10^{12}-10^{14}$~cm$^{-2}$)for \OVI\ (right) identified with the contour method. The center of
    the absorber is defined as the lixel with the highest number
    density. The dashed line indicates a power-law fit of the
    average number density.  We adopt cutoff densities of
    $10^{-12}$~cm$^{-3}$ (\HI) and $10^{-13}$~cm$^{-3}$ (\OVI). }
  \label{fig:spatial}
\end{figure*}

To check that the characteristic density approach is reasonable
at all we examine Figure
\ref{fig:spatial}, where we average the number densities for a set
of absorbers with low column densities ($N_{HI}=10^{12}-10^{13}$~cm$^{-2}$,
$N_{OVI}=10^{13}-10^{13.5}$~cm$^{-2}$) over physical distance along
the line of sight from the peak
number density for the region. A total of 5692 and 1319 absorbers were
used for \lya\ and \OVI\ respectively. The solid line indicates
the average number density, and the shaded region shows $1\sigma$
deviations. The averages were taken by creating bins of physical
distance and, for every absorber, assigning the lixel with the most
closely corresponding physical distance from peak number density to
the appropriate bin. We then fit the resulting distribution with a
power law at small radii and a constant at larger radii (i.e., we fit 
$n(r) = r^\alpha + b$ for $r<d$ and $n(r)=n_o$ for $r>d$). The
slope of the power law, $\alpha$,  allows us to
assess the extent to which an absorber will be dominated by its peak, and
correspondingly, how much a change in characteristic
density fraction could affect column
density weighted properties for absorbers of a given species. The
constant fit for the tail of the distribution, $b$, reflects the average ion 
number density of the regions \textit{between} absorbers. This number
density is adopted as the mean number density of the species, but again 
this mean refers to the mean of the background between absorbers, not
the mean number density of the entire volume or a given sightline. We note
that the constant background value is a very low \OVI\ number
density.  Examination of the raw simulation data shows that these values correspond to cells in the
simulation volume with baryon densities of $10^{-3} - 10^{-1}$ of the cosmic
mean, metallicities of $10^{-4} - 10$~Z$_\odot$, and temperatures of
$10^5$~K or below (resulting in \OVI\ fractions of $10^{-7}$ or
below at the densities in question when the metagalactic ultraviolet
background is taken into consideration).

%The number densities we observe at small distances in Figure~\ref{fig:structure} are comparable to a back of the envelope
%calculation we use to estimate the \HI\ and \OVI\ number densities we
%expect to see for absorbing structures in the WHIM. For \HI\, we assume
%nearly fully photoionized equilibrium in a low redshift radiation field.
%This gives us $n_{\mathrm{HI}}=n_en_H(\alpha_H/\Gamma_H)=(12.8\,
%\mathrm{cm}^{-3})(\mathrm{n}_H)^2$ for a photoionization rate
%$\Gamma_H=2.28\times10^{-14}\,\mathrm{s}^{-1}$, case-A
%radiative recombination rate coefficient
%$\alpha_H=2.5\times10^{-13}\,\mathrm{cm}^3~\mathrm{s}^{-1}$, and an electron density
%$n_e=1.165\mathrm{n}_H$. We expect a hydrogen overdensity factor of at least 1 for a
%given absorber, so using this in conjunction with a mean $n_H$ estimate of
%$n_H=[\Omega_b(1-Y)\rho_{cr}]/m_H\simeq2\times10^{-7}\,\mathrm{cm}^{-3}$, we arrive
%at a rough estimate of $n_{HI}=10^{-12.3}\,\mathrm{cm}^{-3}$.

%To relate the \OVI\ number density to the number density of \HI, we use an
%expected metallicity of $0.1 Z_\odot$, a Solar ratio of
%$(O/H)_\odot=4.9\times10^{-4}$, and an estimate of the ratio
%of $f_{\mathrm{OVI}}$ to $f_{\mathrm{HI}}$ in the WHIM ranging from
%$10^4 - 10^5$ ($5.3 \leq \log T \leq 5.7$)
%to arrive at an estimate of $n_{\mathrm{OVI}}=f_{\mathrm{OVI}}n_O= 
%n_{\mathrm{HI}}(O/H)_\odot(Z/Z_\odot)(f_{OVI}/f_{HI})\simeq
%10^{-12.6}$~cm$^{-3}$ or higher. 

We recognize that due to the shallowness of the power law for \HI\
number density ($\alpha_{HI} = -2.4$), varying the characteristic density 
fraction may affect the properties of the absorbers, especially if a very low
characteristic density fraction is chosen. In this case, contour-identified absorbers will fail to
reflect actual structures, and quantities that should be dominated by peak values
will begin to deviate due to the large excess of sampled
material. \OVI\ should be much less sensitive to the choice in cutoff
given the steepness of the power law ($\alpha_{OVI} = -6.3$).

\begin{figure*}[ht!]
\begin{center}
    \includegraphics[width=.9\textwidth]{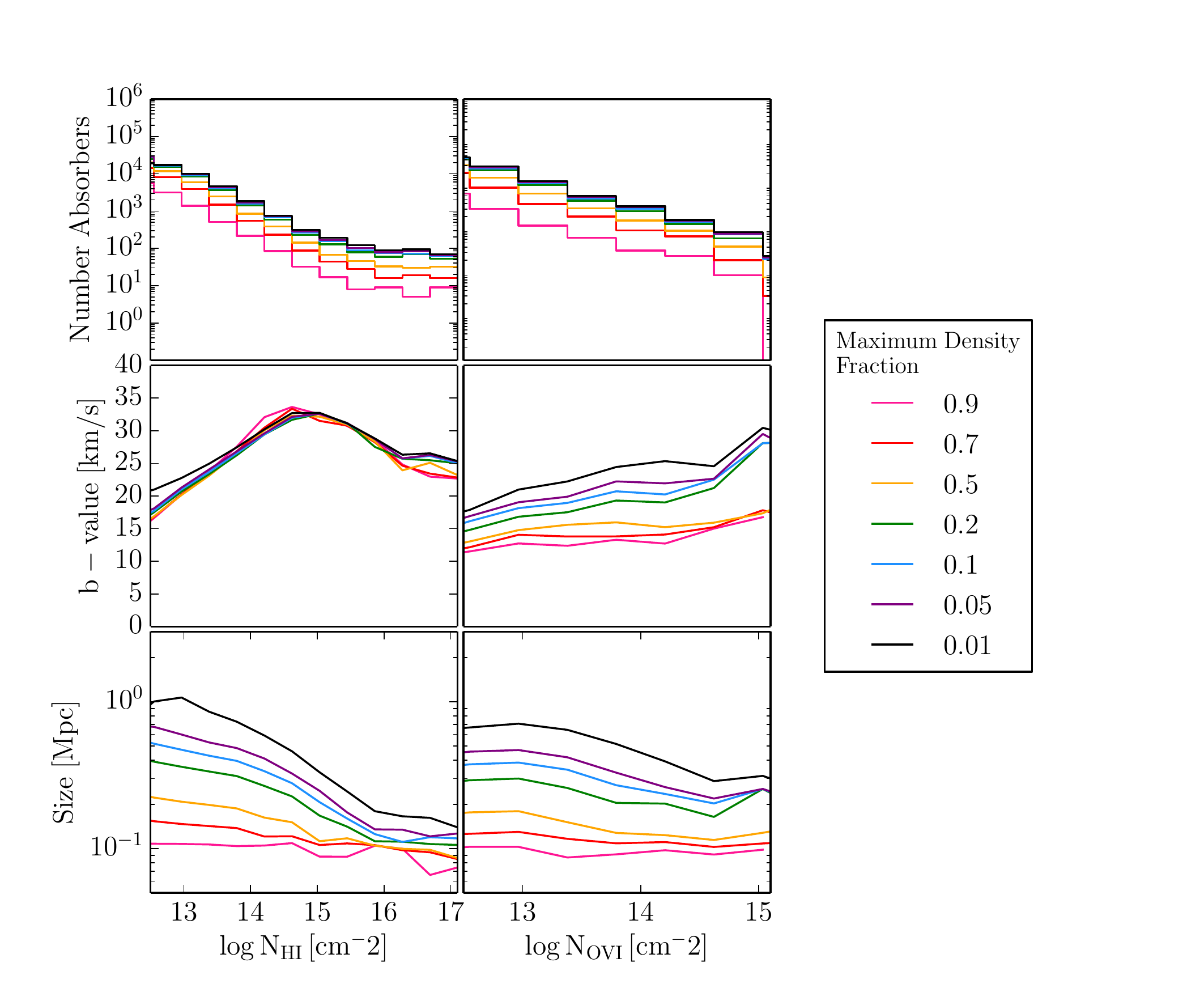}
  \caption{ Absorbers were identified using the contour method for a variety of characteristic density fractions (\emph{colors}). For each set of identified absorbers number of absorbers (\emph{Top}), average b-value (\emph{Middle}), and average size (\emph{Bottom}) were found as a function of column density. \emph{Left:} \HI\ absorbers, \emph{Right: } \OVI\ absorbers.}
  \label{fig:char_frac_tests}
\end{center}
\end{figure*}

This intuition is reflected in Figure \ref{fig:char_frac_tests}, where we show
number of absorbers, line of sight size, and $b$ as a function of 
column density for a variety of characteristic density fractions.
For each cutoff, a corresponding set of absorbers is found independently. 
This analysis is performed for a subset of the total light rays; however,
the results are representative of the total sample. We find as expected that
most quantities are fairly robust to changing characteristic 
density fraction, as displayed by 
the consistency of the b-value results.  The linear size of absorbers, which is not weighted by number
density, increases dramatically when lowering the characteristic density
fraction.  We further investigate the size as a function of column density in 
section \ref{sec:size}. Given the later results we chose a
  characteristic density that is 0.5 times the absorber's maximum baryon density.

\subsubsection{Absorber Size}
\label{sec:size}

\begin{figure*}
\begin{center}
    \includegraphics{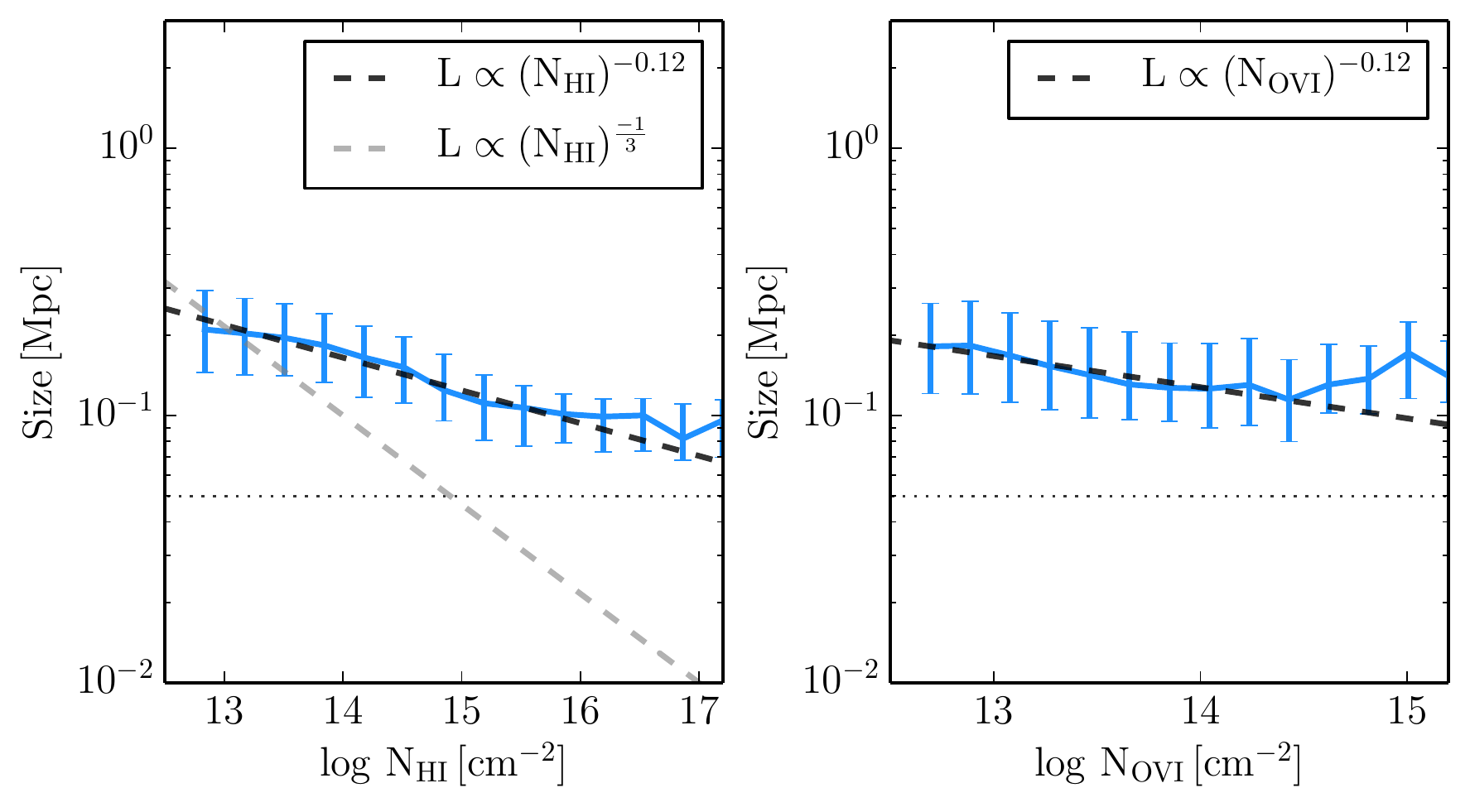}
  \caption{Median size of absorbers binned by column density using . Error bars 
    given by 1st and 4th quartiles. Black dashed line shows line of best 
    fit for median values weighted by number of absorbers in each bin, 
    with fit given in the legend. Grey dashed line in left hand panel shows 
    expected analytic relation, normalized to a 100~kpc absorber at 
    a column density $N_{\mathrm{HI}}=10^{14}$~cm$^{-2}$. 
    Dotted line indicates the resolution limit of the simulation.
    \emph{Left:} \HI\ absorbers. \emph{Right:} \OVI\
    absorbers.}
  \label{fig:size_Var}
\end{center}
\end{figure*}

As absorber size was the most clearly varying parameter with characteristic 
density fraction, it is a useful metric to assess the choice of said 
fraction. Analytic estimates \citep{2001ApJ...559..507S} give an expected 
size-column density relation of $L\propto N^{-1/3}$ for \HI\ absorbers. 
In Figure \ref{fig:size_Var} we plot mean absorber size as a function of 
column density for \HI\ and \OVI. We also overplot a line of best fit for 
the size-column relation obtained through a least squares fit of the median 
value in each column density bin weighted by the number of absorbers in the 
bin. We find a best-fit power law of $-0.12$ for \HI, which is substantially 
shallower than the expected slope of $-1/3$.
%however, the assymetric error 
%bars giving the first and fourth quartiles indicate that a steeper slope 
%could also be consistent with the data. 
For \OVI\ we find a power law of $-0.12$.
%These trends both begin to flatten out once sizes reach 
%$\sim 0.1$~Mpc, which is approaching the simulation resolution, so we only 
%fit the power law to absorber bins with column densities below 
%$10^{16}$~cm$^{-2}$ and $10^{14.5}$~cm$^{-2}$ for \HI\ and \OVI\ respectively.

\subsection{Comparison of Methods}

\subsubsection{Comparison of Contour and Cut Methods}
\label{sec:cutcontour}
Although we do not expect \textit{a priori} that the cut and contour methods 
will give identical results,  we give a short comparison of these
methods. It is important to show comparison to the cut method despite
its lack of physical motivation, as it was used
for a previous analysis of these simulations in \citet{britton2011}. In
order to correlate the absorbers, each absorber from the contour
method is matched by redshift with an absorber from the cut method.

Matching absorbers between methods is accomplished by finding an
absorber from the contour method within a given redshift ($\delta z =
10^{-6}$) of an absorber from the cut method. This process is then
repeated several times after increasing the tolerance by an order of
magnitude until $\delta z = 10^{-3}$. As the redshift range we consider ($0\leq
z\leq 0.4$) is quite small in comparison to the range of redshift windows we
consider, we do not bother to use a window of size $\delta z/(1+z)$. We have chosen to match the absorbers
in this way to create a robust way of matching as many absorbers as
possible, while still ensuring that the matches are as accurate as
possible. Changing the start and end tolerances by an order of
magnitude has minimal effect on the final matches, as long as the
matching is accomplished using a series of monotonically increasing
window sizes.

\begin{figure*}
\begin{center}
  \includegraphics[width=.95\textwidth]{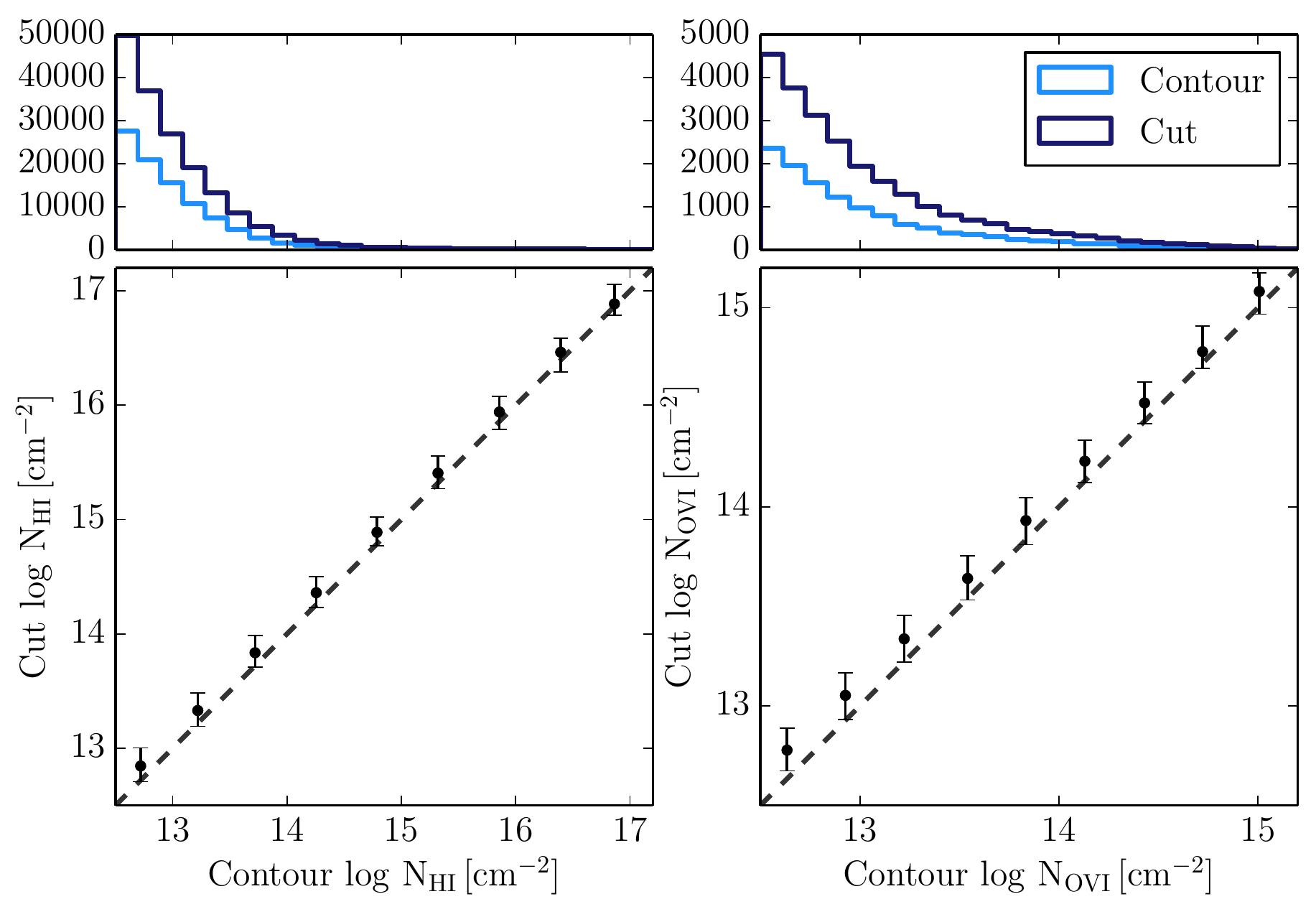}
  \caption{\emph{Top: } Total number of absorbers found in each column
density range for contour method (\emph{light blue}) and cut method
(\emph{dark blue}). \emph{Bottom: } The contour absorbers were binned by
column density and the median column density of the matching cut
absorbers was plotted against the median column density of the
contour absorbers for that bin with vertical error bars showing
first and fourth quartiles in cut column density. Dashed line
shows $\rm{N}_{\rm{contour}}=\rm{N}_{\rm{cut}}$.  \emph{Left: } \HI\
absorbers. \emph{Right: } \OVI\ absorbers.}
  \label{fig:NVar_cutcontour}
\end{center}
\end{figure*}

The column densities of each method for the appropriately matched
and then binned by contour column density. Median contour and cut
column densities and first and fourth quartiles were found for
each bin. This comparison, shown in the lower panels of Figure
\ref{fig:NVar_cutcontour}, indicates no obvious systematic difference 
at any column density. This is indicative of the absorbers being dominated
by the sharp peaks in number density.  

Additionally, the total number of \lya\ absorbers found at low column
densities, shown in the upper panel of Figure \ref{fig:NVar_cutcontour}, is
slightly larger when using the cut method. This is evidence of
the splitting effect, as more absorbers of low column will be found
when a single physical feature is inappropriately identified as two
absorbers.

Given that these results show a comparable column density result, the
contour method is clearly an adequate substitute in analyzing the
direct output. Furthermore, since the contour method finds fewer total
absorbers than the cut method (as it effectively combines multiple
artificially-segmented absorbers into a single, physically meaningful object), as well as absorbers that directly
correlate to the actual cosmological structure in the simulation
volume, it is straightforward to correlate absorbers found with the
contour method with those found using the spectral method. Thus, all
of the analysis in the following sections of this paper will be
completed using values found through the contour method.

\subsubsection{Comparison of Contour and Spectral Methods}
\label{sec:NVar}

We perform an analysis similar to Section~\ref{sec:cutcontour} in
comparing the contour and spectral methods. After matching absorbers
from the contour method with the column density determined by
a single component line fit using the spectral method technique,
the contour absorbers were then binned by
column density. The median spectral and contour column densities for
the absorbers in each of these bins was then plotted with error
bars showing the corresponding first and fourth quartiles for each
bin. These matches were determined using the same
increasing redshift tolerance process as described in
Section~\ref{sec:cutcontour}. 
        
\begin{figure*}
\begin{center}
    \includegraphics[width=.95\textwidth]{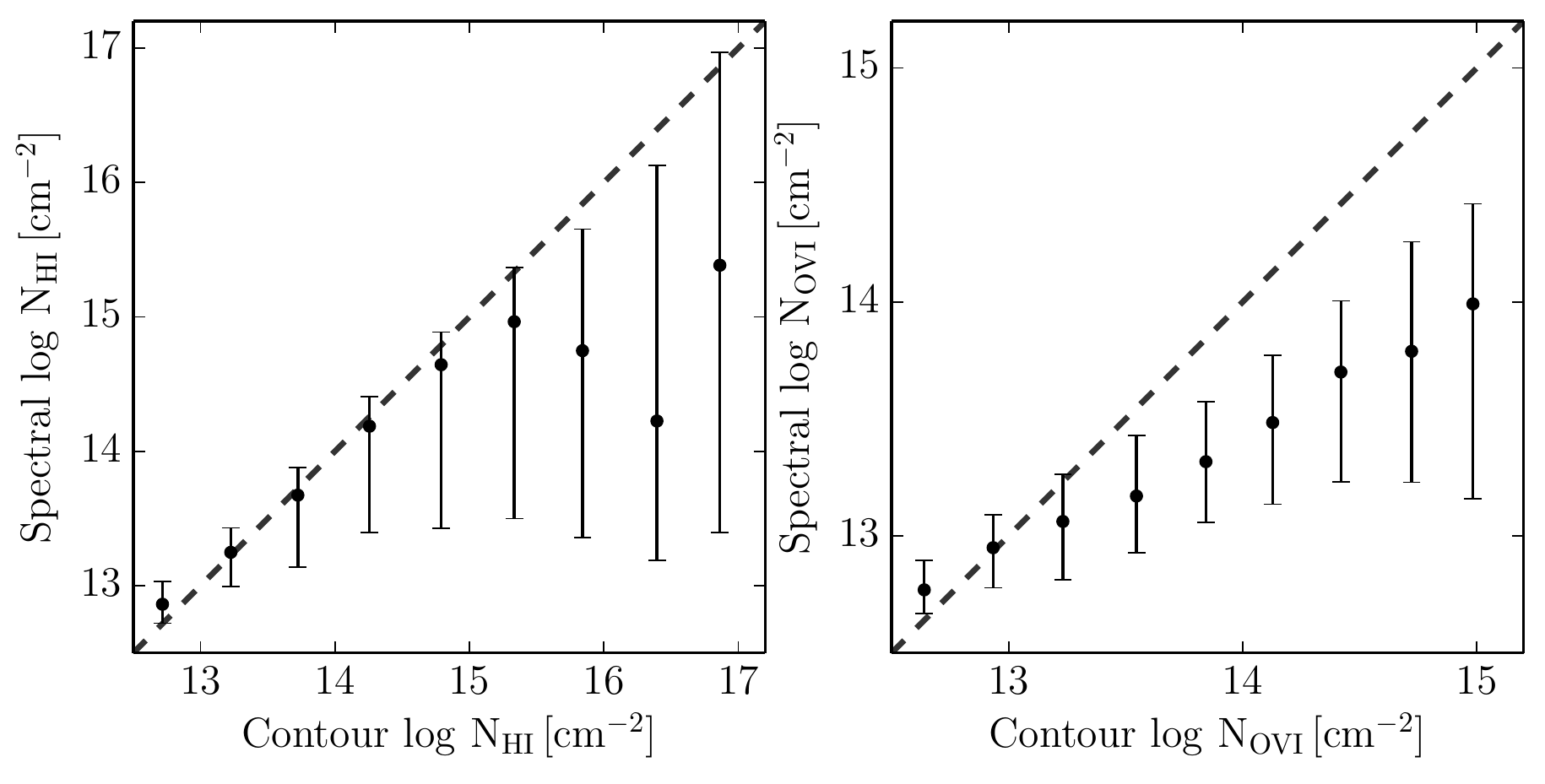}
  \caption{ The contour
    absorbers were binned by column density and the median column
    density  of the matching spectral absorbers was plotted against
    the median column density of the contour absorbers for that bin
    with vertical error bars showing first and fourth quartiles in spectral
    column density and horizontal error bars showing first and fourth 
    quartiles in contour column density. Dashed line shows
    $\rm{N}_{\rm{contour}}=\rm{N}_{\rm{spectral}}$. The spectral method 
    treats each component found in multi-component fits of
    a line complex separately.  \emph{Left: } \HI\
    absorbers. \emph{Right: } \OVI\ absorbers. }
  \label{fig:NVar_split}
\end{center}
\end{figure*}
        
We find that for \lya\ the pipeline has significant trouble
appropriately matching corresponding absorbers at all column
densities and significantly underpredicts the results from the contour method.
For \OVI\ the contour method finds higher column densities
by approximately a factor of $2$ for low column densities
($N_{OVI}=10^{12}-10^{14}$~cm$^{-2}$) and increasing to an order of
magnitude for higher column densities
($N_{OVI}=10^{14}-10^{16}$~cm$^{-2}$).

This is expected as the spectral method is not biased against using
multiple component fits if the $\chi^2$ error is lower in using more 
components. In order to negate this effect, we sum together components
identified in the same complex. A complex is defined as a contiguous region 
with a flux less than $F/F_{continuum}=0.99$. The total column density 
of the region is thus $N_{total} = \sum N_i$. The redshift of the
complex is slightly more difficult to define, but we use a column
density weighted average to assign a single value of redshift to 
the entire complex. Performing the analysis in this way loses some
information about physical structure from line of sight velocities 
that the spectral method is able to identify, but allows us to check 
if the two methods track similar amounts of total material

% Lines that require multiple
%components, with a broad line of
%low column density and a narrower line of higher column density are in
%fact quite common in our simulated \HI\ absorption line spectra, and
%have a physical equivalent -- the Broad \lya\ Absorber, or BLA.  BLAs
%are been seen quite frequently in high signal-to-noise COS and STIS
%spectra \citep{2010ApJ...710..613D}, and are characterized by gas with
%low \HI\ fraction and high ($\geq 10^5$~K) temperatures.  

\begin{figure*}  
\begin{center}
    \includegraphics[width=.95\textwidth]{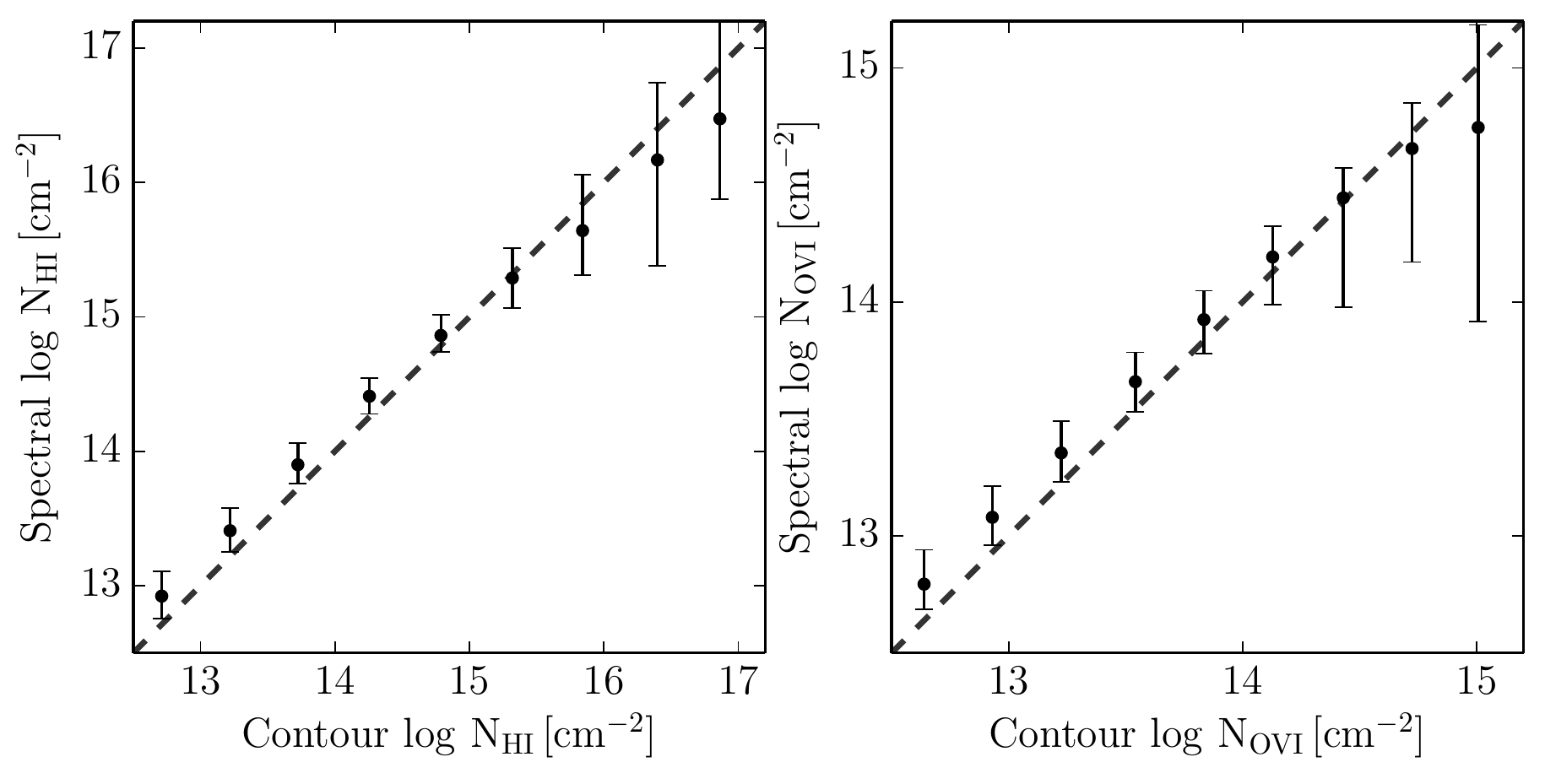}
  \caption{The contour
    absorbers were binned by column density and the median column
    density  of the matching spectral absorbers was plotted against
    the median column density of the contour absorbers for that bin
    with vertical error bars showing first and fourth quartiles in spectral
    column density. Dashed line shows
    $\rm{N}_{\rm{contour}}=\rm{N}_{\rm{spectral}}$. The spectral method 
    combines each component found in multi-component fits of
    a line complex into a single absorber whose column density is the sum of
    the components' column densities.   \emph{Left: } \HI\
    absorbers. \emph{Right: } \OVI\ absorbers.} 
  \label{fig:NVar_combine}
\end{center}
\end{figure*}

As seen in Figure \ref{fig:NVar_combine}, matching the contour method
absorbers to a line complex rather than a line component
gives more comparable column density results for \lya\ and \OVI\ at
all column densities.
There is a larger scatter about the mean for absorbers with higher
column densities ($\mathrm{N_{HI}}\geq 10^{16}$~cm$^{-2}$,
$\mathrm{N_{OVI}}\geq 10^{14.5}$~cm$^{-2}$). This may be indicative
of the difficulty of fitting saturated lines. 
  We can now be reasonably certain that the spectral
and contour methods identify the same absorbing structures, and we can
match them reasonably effectively. The spectral method may find different underlying
IGM substructure than the contour method due to line of sight velocity effects, but the two methods find roughly
equivalent bulk material.

\begin{figure*}  
\begin{center}
    \includegraphics[width=0.95\textwidth]{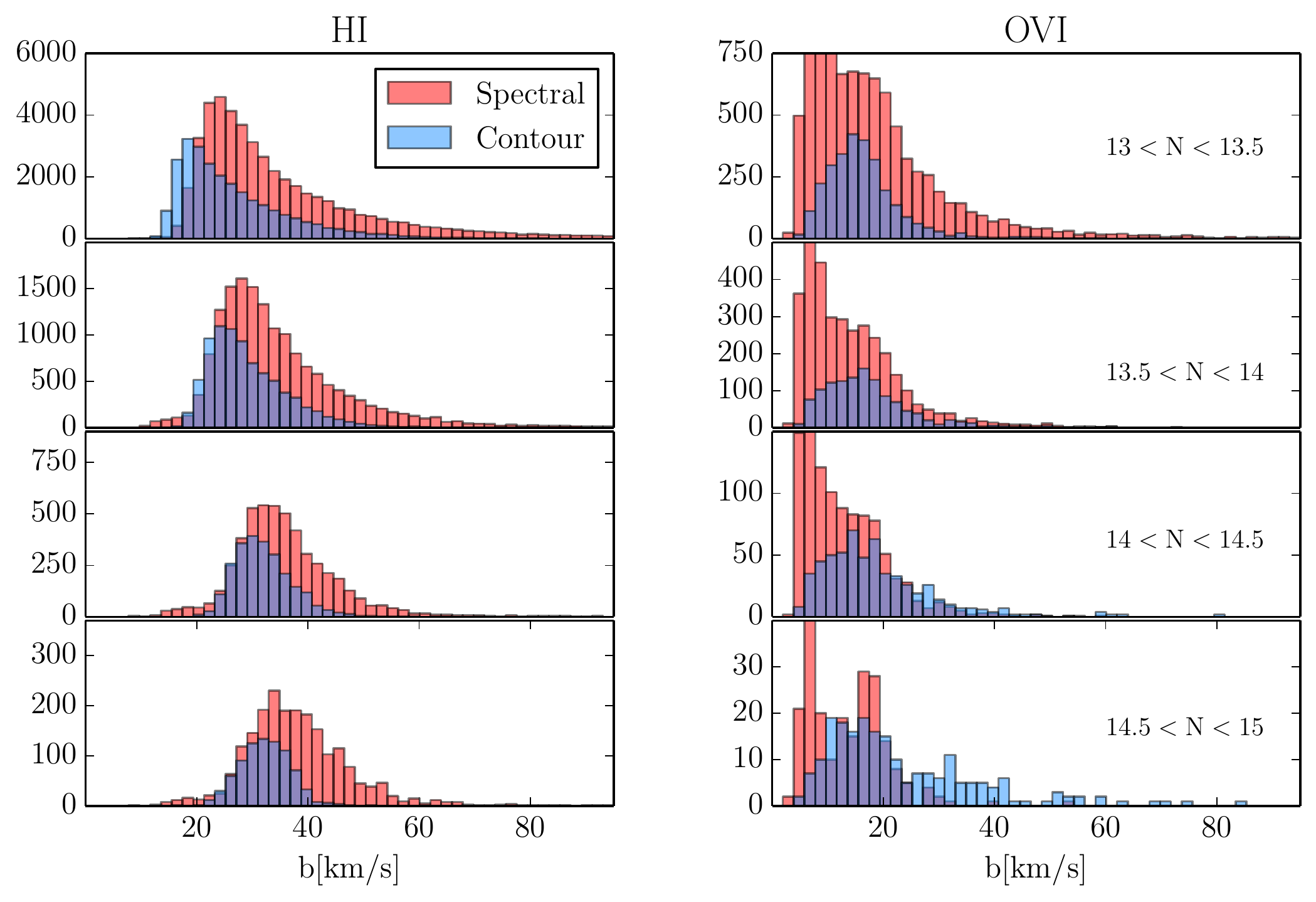}
  \caption{ Histogram of b-values for absorbers with column densities in ranges $10^{13}<N<10^{13.5}$~cm$^{-2}$ (\emph{Top row}), $10^{13.5}<N<10^{14}$~cm$^{-2}$ (\emph{2nd row}), $10^{14}<N<10^{14.5}$~cm$^{-2}$ (\emph{3rd row}), and  $10^{14.5}<N<10^{15}$~cm$^{-2}$ (\emph{Bottom row}). Red bars indicate spectral absorbers, blue bars indicate contour absorbers, and purple indicates an overlap in the distributions. \emph{Left: } \HI\ absorbers. \emph{Right: } \OVI\ absorbers.} 
  \label{fig:sliced_b_distribution}
\end{center}
\end{figure*}

In Figure \ref{fig:sliced_b_distribution} we compare the distribution of b-values for \HI\ and \OVI\ absorbers over slices in column density. The slices were chosen to give a relative idea of of how the methods compare for absorbers of interest; as such, we do not show higher column density slices for \HI, despite the fact that higher column absorbers are fit. We find that for \HI, the distribution of b-values is shifted for the spectral method relative to the contour method at low column ($10^{13}\leq N \leq 10^{13.5}$~cm$^{-2}$) with the peak of the spectral distribution at 25~km~s$^{-1}$ compared to 20~km~s$^{-1}$ for the contour method. For increasing column density slices this distinction shifts with spectral the spectral distribution peaking around 30~km~s$^{-1}$ and the contour distribution peaking around 25~km~s$^{-1}$. The \OVI\ distributions look quite distinct for all column density slices. It seems that compared to the contour method, the spectral method identifies an overabundance of low column absorbers with very low b-values as well as a relative lack of high column aborbers with higher b-values. 
        
\subsection{Comparison to Observations}
We now compare our synthetic absorber population with observed absorbers to assess any systematic differences, attributable to our simulation or methods. 

\subsubsection{Differential and Integral dN/dz}

A common test of the accuracy of simulations is to recreate the
observed number density of \OVI\ absorbers per unit redshift
\citep{tilton2012,danforth2008}, as well as the cumulative number
density of \OVI\ absorbers per unit redshift
\citep[e.g.,][]{fang2001,cen2006,oppenheimer2009,teppergarcia2011,britton2011}.
It is thus quite useful to understand how generating this statistic
from simulations using an observationally-motivated method
intrinsically differs from generating this statistic from absorbers
found using analysis of cell-by-cell output.

\begin{figure*}
\begin{center}
  \includegraphics[width=.95\textwidth]{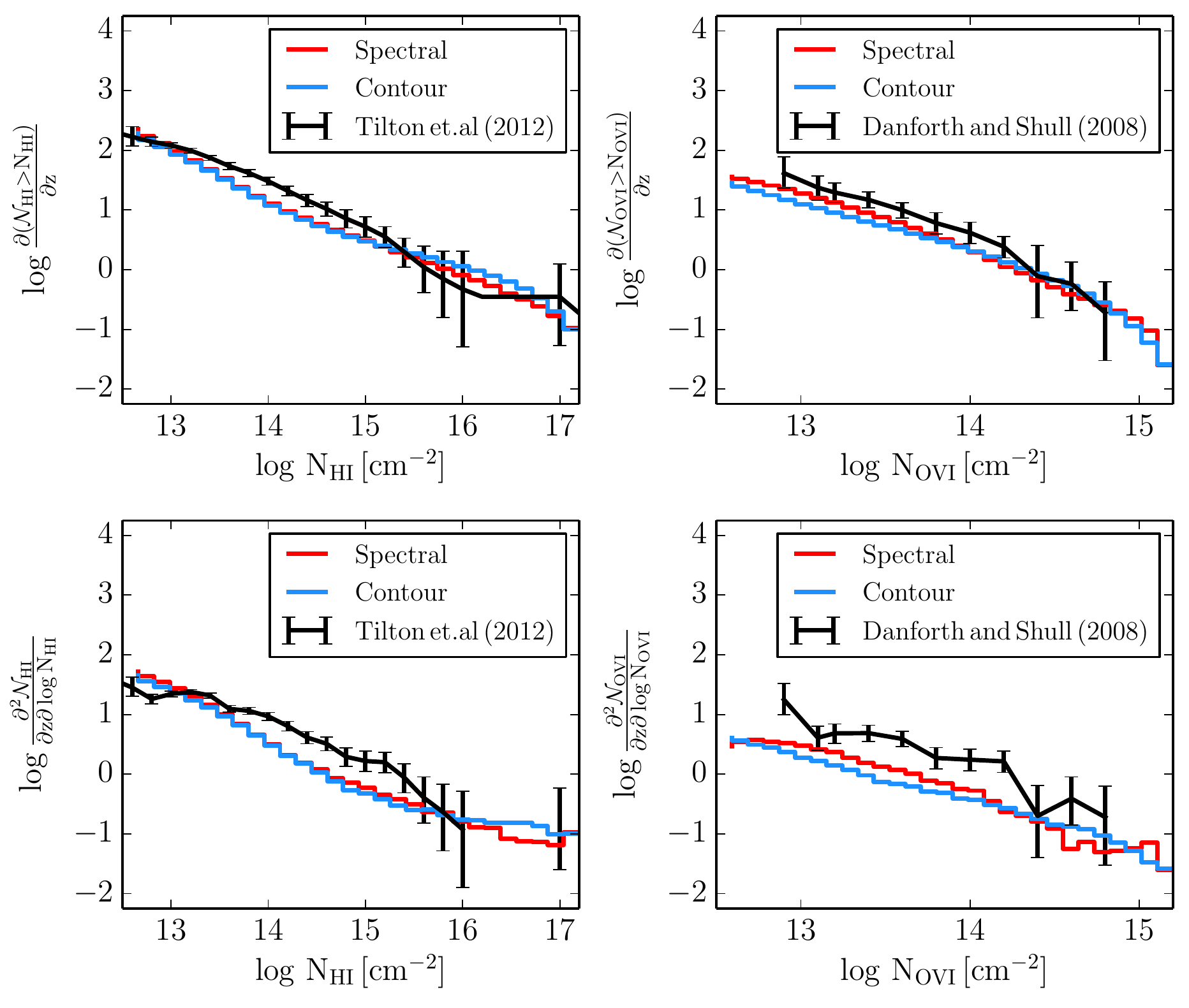}
  \caption{\emph{Top: } Number of
    absorbers ($\mathcal{N}$) with a column density greater than N
    per redshift z (i.e., cumulative line number density). \emph{Bottom:} Number of absorbers ($\mathcal{N}$) per column
    density ($N$) per redshift ($z$) (i.e., line number density). \emph{Left:} \HI\ column densities with
    observational points and error bars from \citet{tilton2012}. \emph{Right:}
    \OVI\ column densities, with observational points and error bars
    from \citet{danforth2008}.  In all panels, the red line
    corresponds to the spectral method, the blue line to the contour
    method, and the black line to observations.}
  \label{fig:dndz}
\end{center}
\end{figure*}

Figure~\ref{fig:dndz} shows this analysis for \HI\ and \OVI\ absorbers
found in our simulation, along
with observational data from \citet{tilton2012} and
\citet{danforth2008} for comparison for \HI\ and \OVI\, respectively. We consider each component
of a multi-component fit of a complex separately because, although
single-component fits were favored in both observational cases,
complexes with clear evidence of substructure were fit in a
multi-component fashion with each component listed separately in the
calculation of $d \mathcal{N}/dz$. In the top row of this Figure we
show the cumulative line number density of \HI\ and \OVI\ shown in the
standard way -- i.e., we show the number of
absorbers above a given species column density at the mean redshift of
the simulation outputs, normalized by the total redshift interval
$\Delta z$ of the synthetic observations.  The bottom row displays the
differential line number density normalized by the total redshift
interval $\Delta z$, also known as the column density distribution function.

The two methods agree quite well at all column densities for both
\lya\ and \OVI\ as seen in both the differential and integral forms.
There is a systematic under-prediction relative to the observed
absorber frequency using either simulated method in \OVI; however, this was also
found in the initial analysis presented by \citet{britton2011}, and is
likely to be an intrinsic property of the simulation rather than a
feature of our method of determining absorber properties.

It should be noted that we consider absorbers above a column
density of $10^{12.5}$~cm$^{-2}$ for the contour method, and fit below
the observable limit for both \HI\ and \OVI\ with the spectral method using noiseless
spectra. As a result, no significant completion correction is required
for low column absorbers in our simulated results, which may result in
systematic differences from the cited observational results.
        
\subsubsection{Broadening Value}

The other parameter that is found directly using the spectral method
is the Doppler parameter, or `b-value.'  This parameter determines the width of a line and
typically has contributions from both thermal and non-thermal broadening.

\begin{figure*}
\begin{center}
    \includegraphics[width=0.95\textwidth]{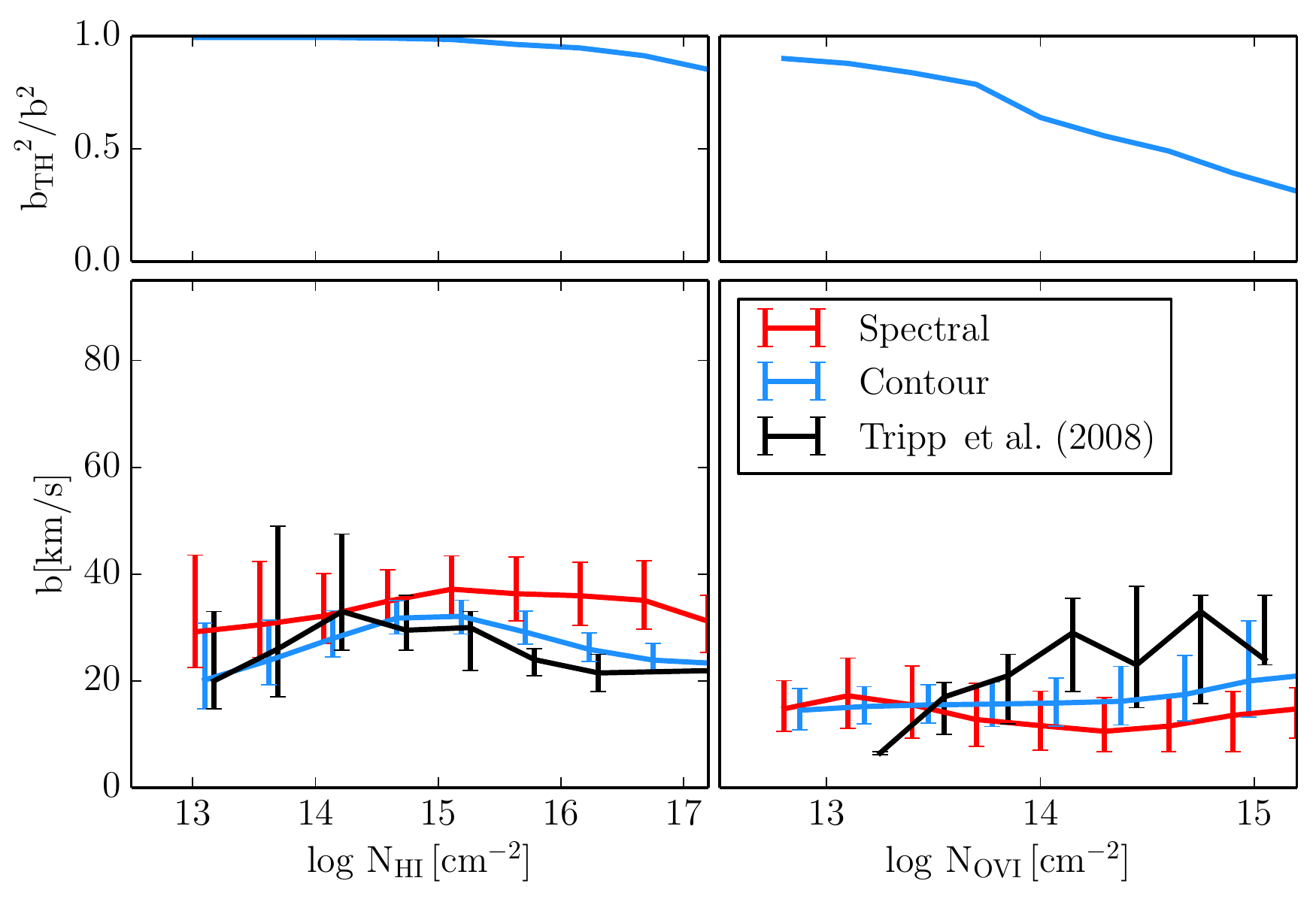}
  \caption{\emph{Top: } Squared fraction (due to component b-value 
    addition in quadrature) of b-value due to thermal motion as a
    function of column density for the contour method. \emph{Bottom: }Mean b-value vs Column
    density for contour (\emph{blue}) and spectral (\emph{red})
    methods. Observational points (\emph{black}) from
    \citet{tilton2012}. Error bars show $1\sigma$
    deviations. \emph{Left:} \HI\ absorbers. \emph{Right:} \OVI\
    absorbers.}
  \label{fig:bVar}
\end{center}
\end{figure*}

While observationally these components are very challenging to distinguish and
require simultaneous fits of multiple species with different masses,
in a simulation we have perfect knowledge
of the thermal and kinematic behavior of the plasma everywhere in our
volume. To that end, a more detailed and precise study is
possible. Figure~\ref{fig:bVar} shows the median b-value plotted against
column density for both the contour and spectral methods, as well as
the squared fraction of total b-value due to thermal motion as a function of
column density for the contour method. We look at the squared fraction as
the thermal and non-thermal components of the total b-value are added in
quadrature.

The \lya\ values agree reasonably well between methods at lower column
densities and but diverge slightly at high column
densities ($N_{\mathrm{HI}}\geq 10^{16}$~cm$^{-2}$). The values increase
slightly with column to $N_{\mathrm{HI}}\sim10^{14.5}$~cm$^{-2}$. The
b-values from the contour method then begin to slightly decrease again
while the values derived from the spectral method remain fairly constant.
Observational results from \citet{tripp2008} show a similar pattern as 
the contour method.

For \OVI, the median b-value increases stays roughly constant at 
$b\sim 20 $~km s$^{-1}$ using the contour method. The spectral method 
stays roughly constant at  $b\sim 15$~km s$^{-1}$. 
%We do not see the trend indicated by \citet{tripp2008} of increasing b-value
%with column density in the spectral method, but there is some indication
%of this trend in the contour method, albiet less dramatically at lower column
%densities.
We do not see the trend indicated by \citet{tripp2008} of increasing
b-value with column density for either method.
We note that the biggest divergence between the two
methods occurs where the thermal component of the b-value becomes less
dominant; however, this does not cause a systematic under or over
prediction of b-value by either method across both species. 

We note that here we only consider the b-values of each individual
spectral component in this analysis. Although combining complexes was
useful in an attempt to correlate spectral structures with contour
structures as in Sections \ref{sec:NVar}, this method obscures the
underlying structure that can be determined with the aid of the
b-values. We thus do not expect the contour and spectral b-value
distributions to be the same, as they simply do not correspond to the
same quantities; the contour method takes the variation in bulk motion
of separate features and gives a higher non-thermal motion component of
the b-value, whereas the spectral method divides up the same region of
physical space into separate absorbing components.

\subsection{Physical Conditions of Absorbers}
In order to extend our understanding of the WHIM and its relationship to observations, we examine the physical environment associated with the absorbers identified with the contour method.

\subsubsection{Median Quantities Of Absorber Systems}

\begin{figure*}
\begin{center}
    \includegraphics[width=.95\textwidth]{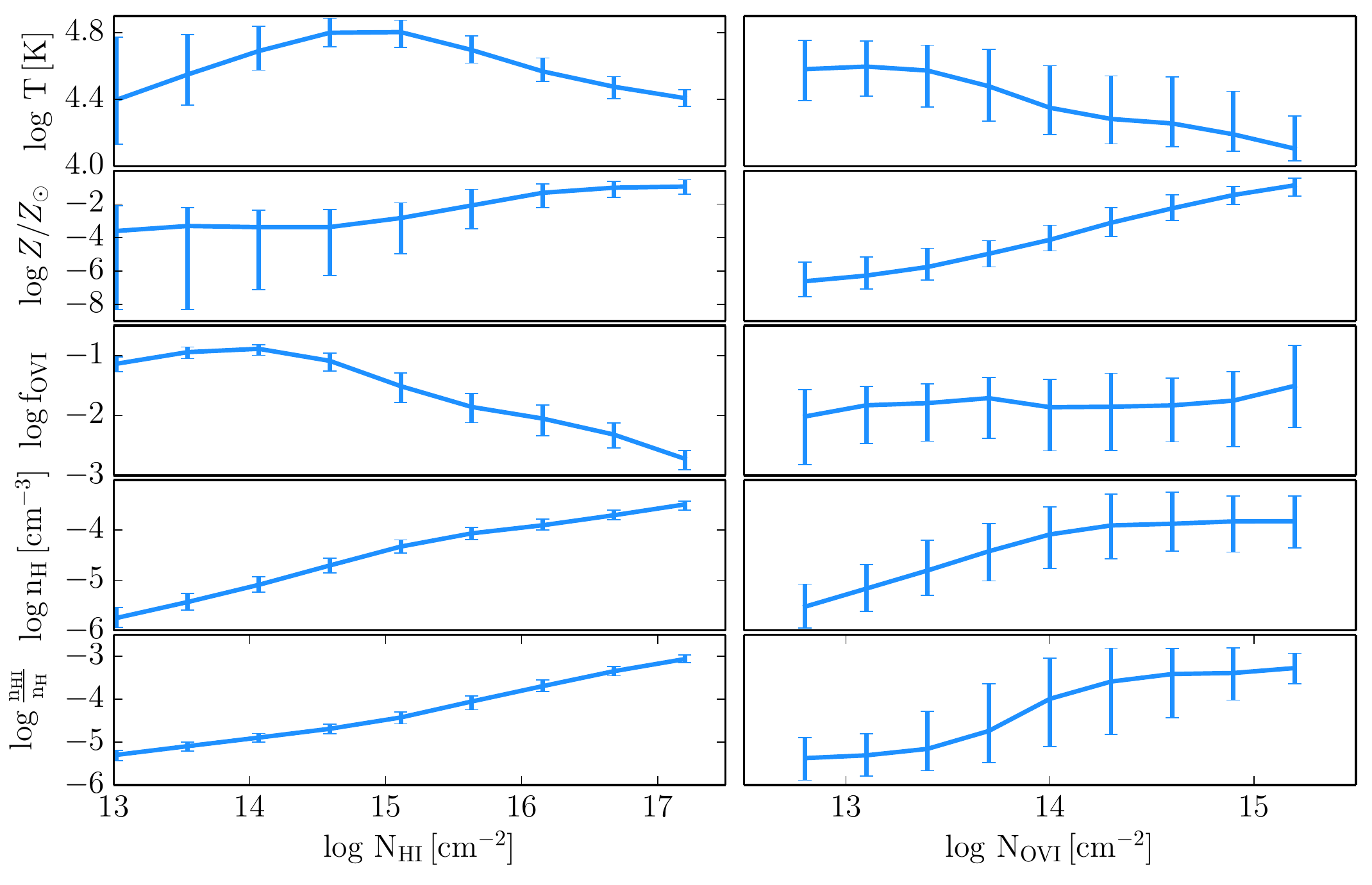}  
  \caption{Median temperature, metallicity, \OVI\ fraction, total H number density,
    and \HI\ fraction plotted against column density for absorbers found using the contour method. Error bars show first and
    fourth quartiles. \emph{Left:} \lya\ absorbers. \emph{Right:} \OVI\
    absorbers.}
  \label{fig:param}
\end{center}
\end{figure*}

Figure~\ref{fig:param} details the physical condition of the gas as a
function of absorber column density. 
Gas temperature appears to increase and then decrease with increasing column
density for \HI\  absorbers, while
the temperature of \OVI\ absorbers decreases with increasing
column. The temperature
behavior for \HI\  also typically has much less variance at a given column density. 
Metallicity, defined as the total mass in elements heavier than helium relative 
to the total gas mass normalized by the Solar metal fraction, increases 
steadily with increasing \OVI\ column density.
There is slight increase with increasing \HI\ column density, albeit 
with an increasing variance at low $\mathrm{N_{HI}}$, suggesting that 
low $\mathrm{N_{HI}}$ systems may trace a large variety of environments. 
Clear trends in \OVI\ fraction 
can be seen as a function of \HI\ column density, with a peak at 
N$_{HI} \simeq 10^{14}$~cm$^{-2}$ and lower fractions at larger and smaller columns.
The median number density of H increases steadily for both \HI\ 
and \OVI\ 
absorbers, although over a wider range with smaller variance for \HI.  The fraction 
of \HI\ ionization steadily increases as a function of increasing \HI column density. The \HI\
fraction also seems to increase with \OVI\ column density, but the variance in this
trend is far larger. These trends appear largely consistent with those initially
provided by \citet{britton2011}. 

\subsubsection{Thermal State of \OVI\ Absorbers}
One of the primary purposes of these simulations was to investigate the 
utility of \OVI\ as a tracer of the WHIM. The initial analysis presented by
\citet{britton2011} showed a bimodality in the temperature distribution
of \OVI\ also seen in \citet{teppergarcia2011}. The bimodality was centered
at $T\sim10^5$ K, with 57\% of \OVI\ absorbers found
around temperatures of $10^{5.5}$ K in the WHIM phase and 37\% found
at temperatures of $10^{4.5}$ K in the warm phase (and the remaining 6\%
found at higher densities in what \citet{britton2011} defined as the
`condensed' phase, with a baryon overdensity of $\Delta_b \geq 1000$). Such a
bimodality suggests that both collisionally ionized and photo-ionized \OVI\ are present in significant amounts in the IGM.

\begin{figure*}
  \begin{center}
  \includegraphics[width=0.85\textwidth]{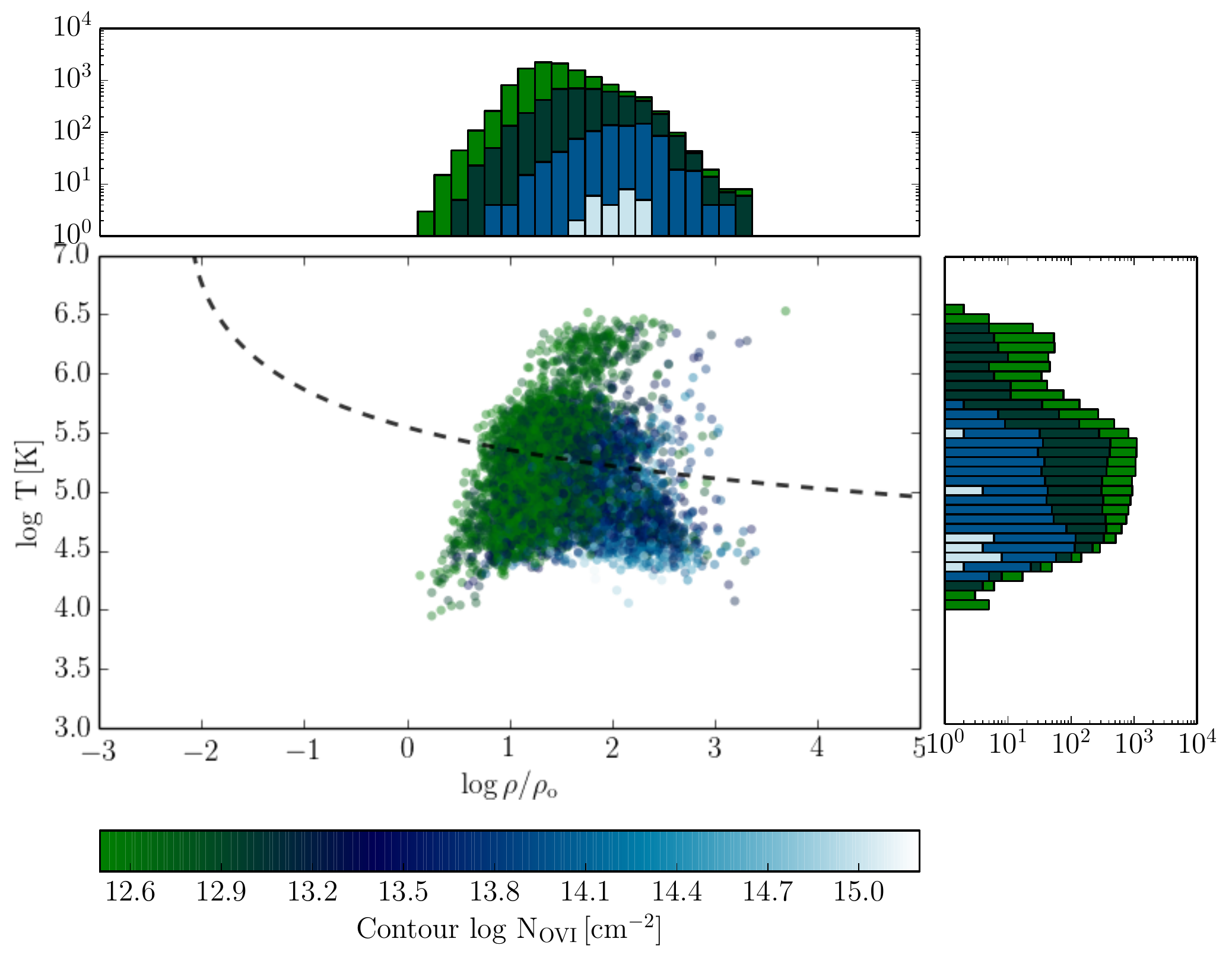}
  \caption{Thermal-state distribution of \OVI\ absorbers.  Each absorber is plotted
    as a function of its mean baryon overdensity
    (defined as $\rho/\rho_o$) and temperature,
    with points colored according to absorber column density. The dashed line
    indicates where collisional ionization equals photoionization (collisional
    ionization dominates above the curve).
    Histograms along the top and right side of the scatter plot show
    overdensity and
    temperature, respectively, for all \OVI\ absorbers with column
    densities greater than $10^{15}$~cm$^{-2}$
    (light blue), 10$^{14.5}$~cm$^{-2}$ (dark blue), $10^{14}$~cm$^{-2}$ (dark green),
     and $10^{13}$~cm$^{-2}$ (light green).}
  \label{fig:phase}
  \end{center}
\end{figure*}

We then must ask if these statistics are proportionally recreated when
looking at absorbers found using the methods presented in this
paper. \citet{britton2011} found there to be no bias between the true phase distribution of \OVI\ in the simulation and that inferred from absorbers created with the cut method,
whereas \citet{teppergarcia2011} found \OVI\ absorption to be biased towards higher temperatures. Figure~\ref{fig:phase} shows the baryon overdensity and
temperature for each absorber found using the spectral method.  Baryon
properties are determined by using the mean overdensity and
temperature of the gas in the same absorber found with the contour method, 
as determined using column density-weighted averages of all cells in the absorber.
The distribution shows no evidence of a strong temperature bimodality, but 
instead shows a roughly smooth distribution over the temperature range 
$4.5 \leq \log (T / K) \leq 5.5$. If a bimodality
does exist it is only present in the very highest column absorbers. There are no
significant discrepancies between the absorbers identified here with
the contour method versus those identified initially with the cut
method. We also find similar overall phase fractions, with $69\%$ of
\OVI\ absorbers in the WHIM phase, $30\%$ of absorbers in the warm
phase, and $1\%$ of absorbers in the condensed phase.
            
\subsubsection{Cool vs. Warm IGM}
\label{sec:temp_bimodality}

\begin{figure*}  

\begin{center}
  \includegraphics[width=0.95\textwidth]{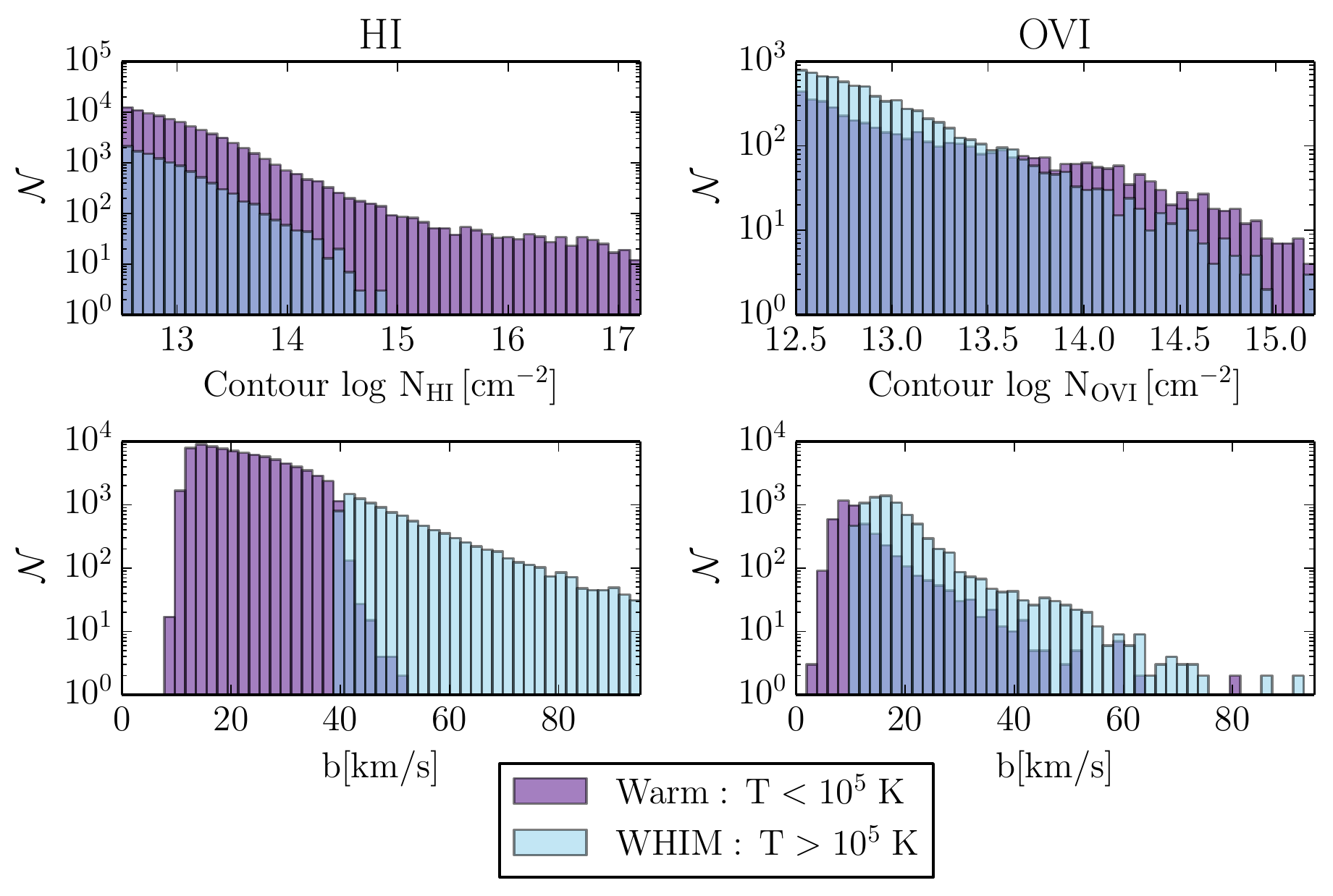}
  \caption{  Absorbers binned by column density (\emph{top}) and b-value (\emph{bottom}). WHIM absorbers are in blue, warm absorbers are in purple.\emph{Left: } \HI\ absorbers. \emph{Right: } \OVI\ absorbers. All absorbers have been identified using the contour method.}
  \label{fig:temperature_bimodality}
\end{center}
\end{figure*}

The WHIM is often defined as gas with temperatures in the range 
$10^5-10^7$ K \citep{1999ApJ...514....1C,2001ApJ...552..473D}. 
Using temperatures derived via the contour method, we attempt to 
determine the characteristics of observables for absorbers in this 
temperature range and assess any systematic differences between 
absorbers with lower temperatures. For this analysis we henceforth
define a WHIM asborber as an absorber with $\mathrm{T>10^5~K}$ and
a warm absorber as one with $\mathrm{T<10^5~K}$.
Figure \ref{fig:temperature_bimodality} 
shows column density and $b$-value histogrammed separately for 
WHIM and warm absorbers. The warm \HI\ absorbers have a roughly 
linear distribution by column density in logspace, while the WHIM 
\HI\ absorbers fall off sharply after $\mathrm{N_{HI}}\sim 10^{14}$~cm$^{-2}$. 
The warm and WHIM \OVI\ absorbers show very similar distributions 
by column density, with the WHIM absorbers dominating slightly below 
$N_{OVI}\sim 10^{14}$~cm$^{-2}$. The $b$-value histograms show 
two distinct distributions for the WHIM and warm absorbers 
for both \HI\ and \OVI. \HI\ warm absorbers have a peak $b$-value 
of $~15$~km~s$^{-1}$ whereas WHIM absorbers peak at around 
45~km~s$^{-1}$. In \OVI, warm absorbers peak at 10~km~s$^{-1}$ while 
WHIM absorbers peak at 20~km~s$^{-1}$.

\begin{figure*}  

\begin{center}
    \includegraphics[width=.95\textwidth]{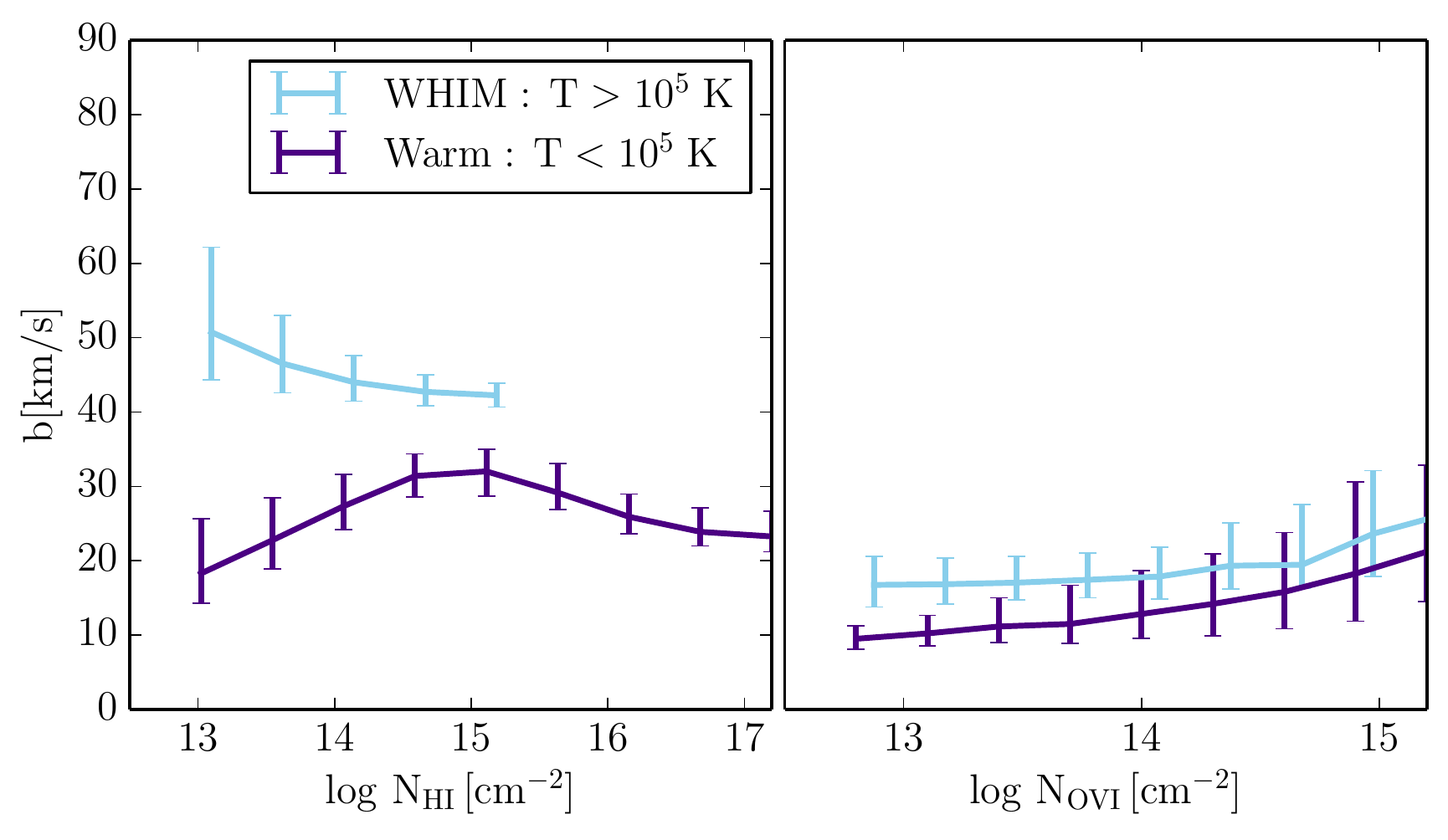}
  \caption{  Median b-value vs column
    density for WHIM (\emph{blue}) and warm (\emph{purple}) 
    absorbers. Error bars show 1st and 4th quartiles. \emph{Left: } \HI\
    absorbers. \emph{Right: } \OVI\ absorbers.} 
  \label{fig:temperature_bVar}
\end{center}
\end{figure*}

After establishing the relative distributions of absorber observables 
in the warm and WHIM phases, we examine the relationship between the 
observables. In Figure \ref{fig:temperature_bVar} we show median 
b-value plotted over column density bins. We find a general decrease 
in the b-value of an absorber with column density for WHIM phase \HI\ 
absorbers. A absorber with column density in the last bin appears to 
reverse the trend, but this point is not statistically significant. 
For \HI\ absorbers in the warm phase, b-values increase with column density 
until roughly N$_{HI}=10^{15}$~cm$^{-2}$ and then begin to decrease again 
slightly. For \OVI\ absorbers both the warm and WHIM absorbers have 
slightly increasing b-values with column density, but the median b-value 
of WHIM absorbers is typically 5-10~km~s$^{-1}$ higher than the median 
b-value of the warm absorbers for a given column density bin.

\begin{figure*}
\begin{center}
    \includegraphics[width=.95\textwidth]{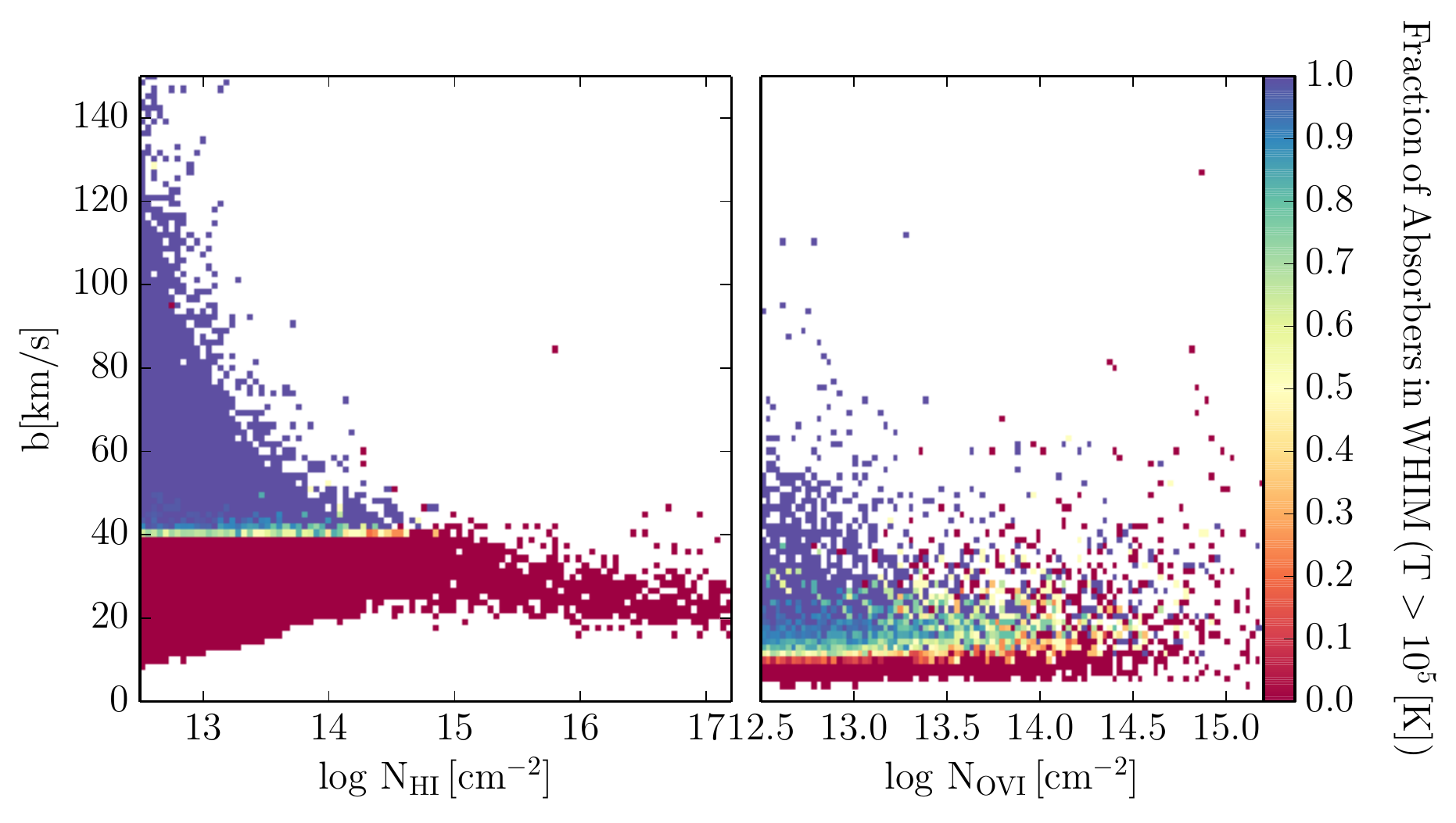}
  \caption{ Fraction of total absorbers that are WHIM absorbers placed into two-dimensional histograms of column density and b-value. The color shows the fraction of total absorbers where red indicates no absorbers in the WHIM and white indicates no absorbers of any phase. \emph{Left: } \HI\ absorbers. \emph{Right: } \OVI\ absorbers.} 
  \label{fig:temperature_2dhist}
\end{center}
\end{figure*}

In Figure \ref{fig:temperature_2dhist} we show the fraction of 
WHIM absorbers out of total absorbers histogrammed two-dimensionally 
along column density and $b$-value. We find that above $b$-values of 
40 and 15 km~s$^{-1}$ for \HI\ and \OVI, respectively, the fraction 
is dominated by WHIM absorbers. These sharp cutoffs are to be expected 
because for a temperature of at least $10^5$ K there is a minimum 
$b$-value given by Equation \ref{eq:bth}. We also see that the higher 
$b$-values are associated with lower column densities, although this 
correlation is stronger for \HI\ absorbers.

\subsubsection{Collisionally Ionized vs. Photoionized \OVI}

In Figure \ref{fig:phase} we plotted a dotted line indicating where 
collisional ionization begins to dominate over photoionization for \OVI, given by 

\begin{equation}
  \begin{split}
    \log \rho / \bar{\rho}_b = 
    7.68  \left( \frac{1+z}{1.2} \right)^{-3}  T^{-1/2} \left(1 + \frac{T}{1.32 \times 10^7}\right)\\
    \times  e^{1.32 \times 10^6 / T}
  \end{split}
\end{equation}

derived in \citet{shull2012}, where $\bar{\rho}_b$ is the
universal mean
baryon density such that $\rho/\bar{\rho}_b = \Delta_b$.  This
relation is valid in the density regime where the IGM is optically
thin to the metagalactic ionizing background, which includes all of
the absorbers shown in Figure~\ref{fig:phase}.  In this section we 
perform a similar analysis as was done in section \ref{sec:temp_bimodality}, 
but instead of making a temperature cut to distinguish between WHIM 
and warm absorbers, we differentiate between absorbers where collisional 
ionization dominates and where photoionization dominates. As this 
differentiation is only appropriate for \OVI, we do not show results 
for \HI\ absorbers here. 

\begin{figure*}  
\begin{center}
  \includegraphics[width=0.95\textwidth]{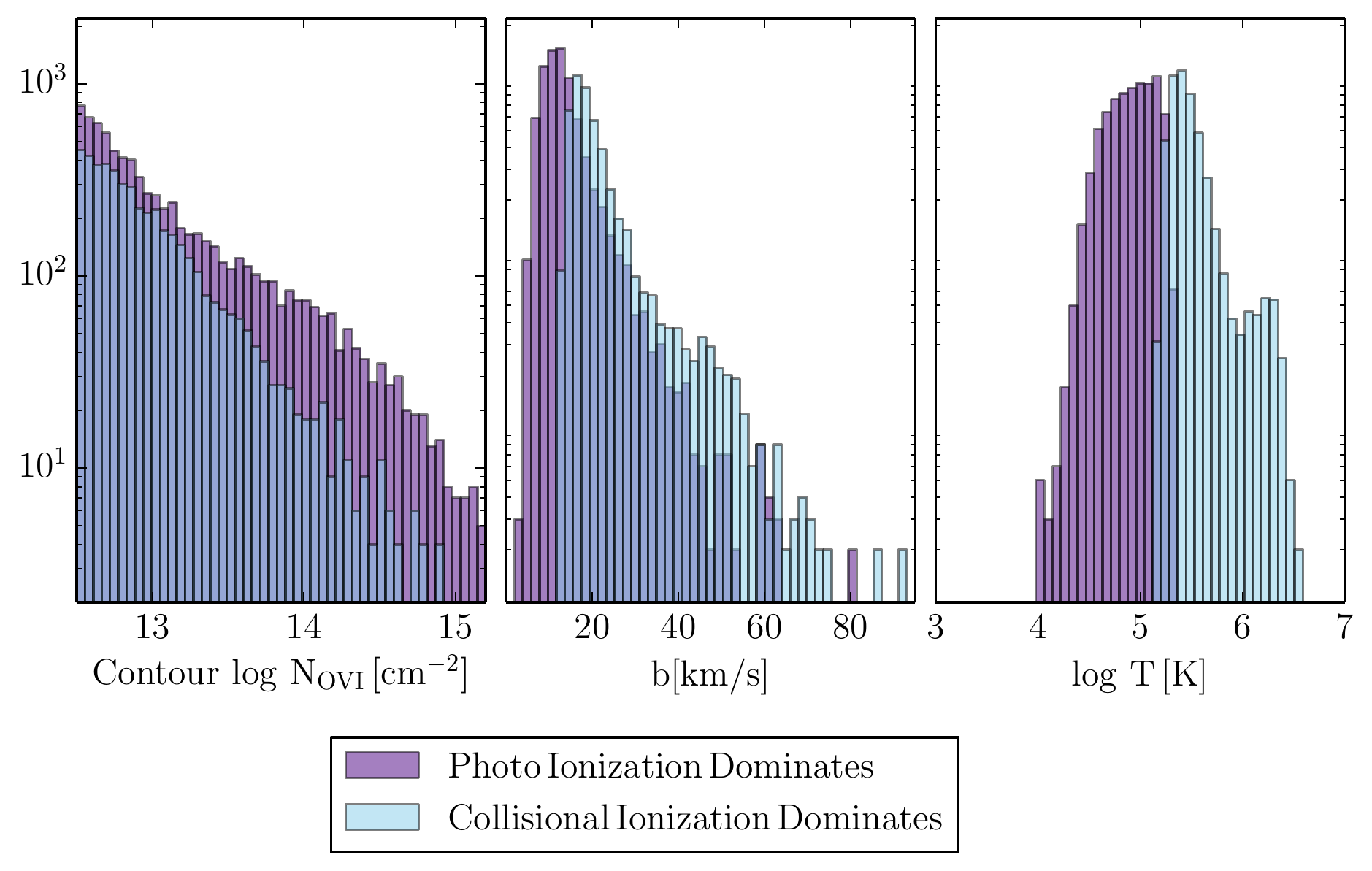}
  \caption{\OVI\ absorbers binned by column density (\emph{left}), b-value (\emph{center}), and temperature (\emph{right}). Collisional ionization dominated absorbers are in blue, photoionization dominated absorbers are in purple. }
  \label{fig:ionization_bimodality}
\end{center}
\end{figure*}

Figure \ref{fig:ionization_bimodality} shows column density, $b$-value, 
and temperature histogrammed for photoionization dominated and 
collisional ionization dominated absorbers. The distribution of 
column density shows no significant difference for the two absorber 
populations. The $b$-value distributions both appear roughly exponential 
but the photoionization dominated population's distribution peaks at 
$b$ $\sim 15$~km~s$^{-1}$. while the collisional ionization dominated 
population peaks at $b\sim 20$~km~s$^{-1}$. Similarly, both temperature 
distributions appear roughly gaussian with photoionization dominated 
peaking at T$\sim 10^5$~K and collisionally ionization dominated at 
T$\sim 10^{5.5}$~K. 

\begin{figure}
\begin{center}
    \includegraphics[width=0.45\textwidth]{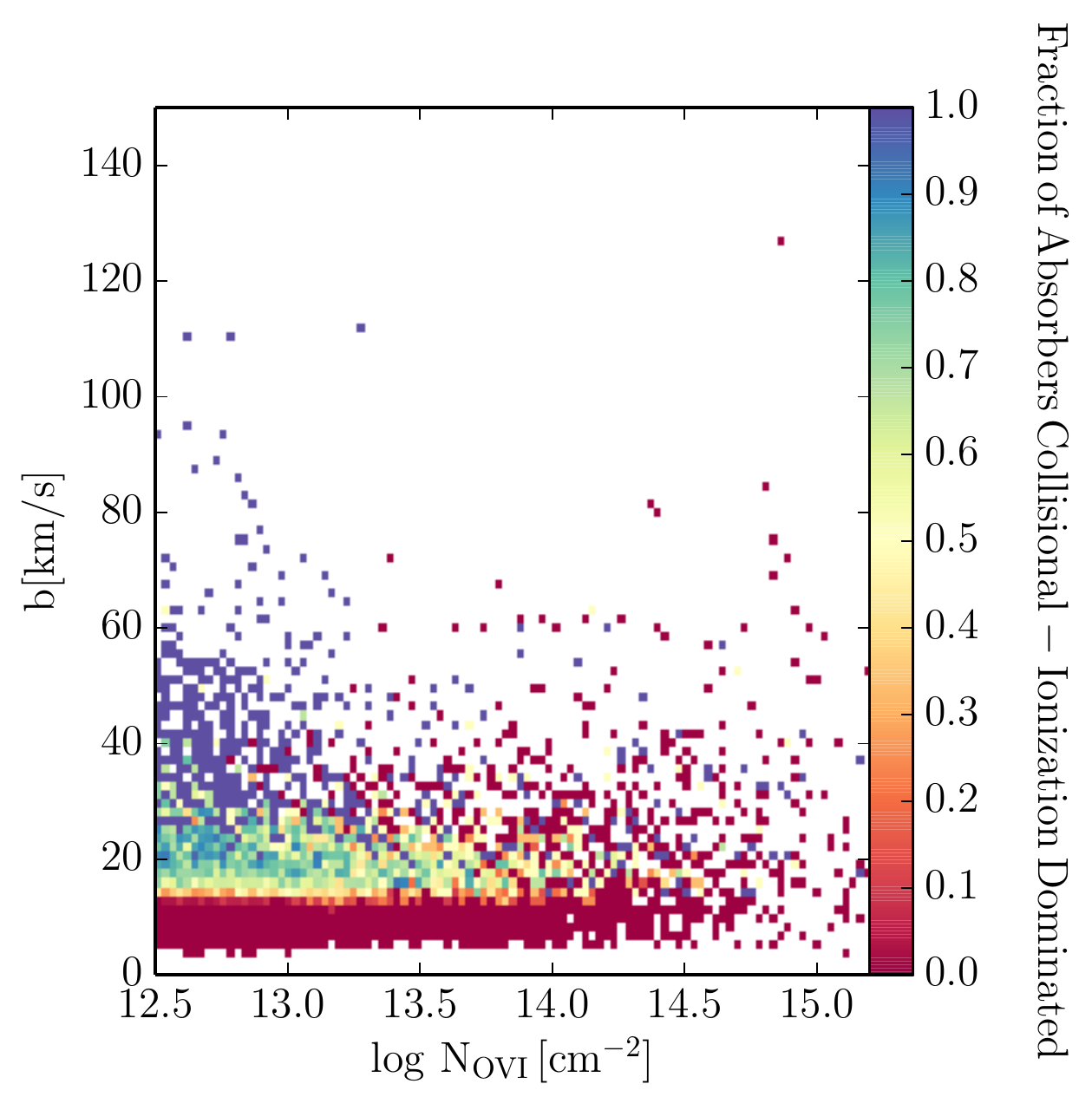}
  \caption{Fraction of total \OVI\ absorbers that are collisional ionization dominated placed into two-dimensional histograms of column density and b-value. The color shows the fraction of total absorbers where red indicates no absorbers in the WHIM and white indicates no absorbers of any phase.  } 
  \label{fig:ionization_2dhist}
\end{center}
\end{figure}

In an approach similar to the one in Figure \ref{fig:temperature_2dhist}, 
Figure \ref{fig:ionization_2dhist} shows the fraction of collisional 
ionization-dominated absorbers out of total absorbers in a two 
dimensional histogram of column density and $b$-value. The results are 
qualitatively similar to those seen in the \OVI\ panel of Figure 
\ref{fig:temperature_2dhist}, but with a slightly smoother transition 
along the $b$-value axis.

\subsubsection{Relating \HI\ and \OVI}

\begin{figure*}  
\begin{center}
  \includegraphics[width=0.95\textwidth]{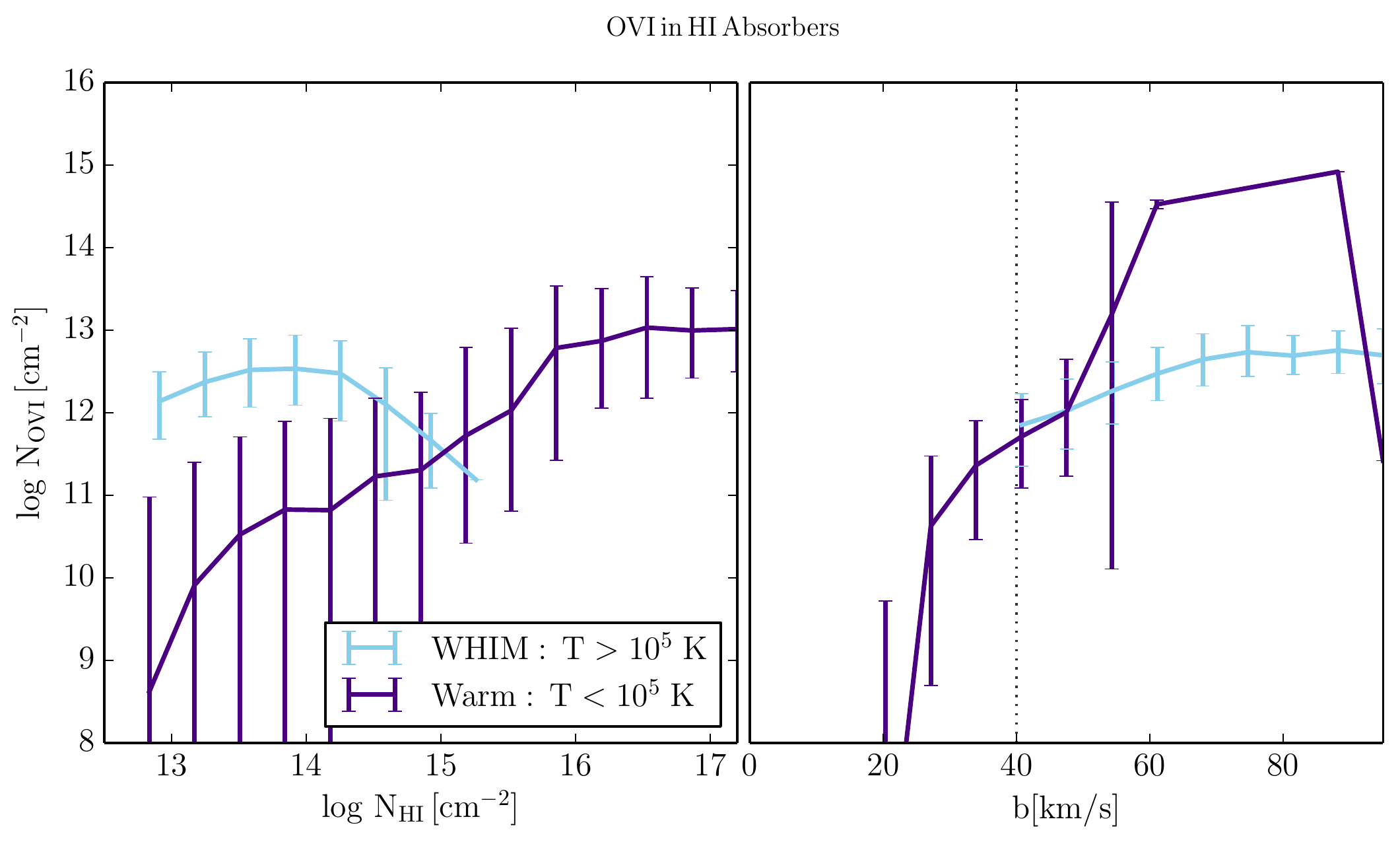}
  \caption{  Median associated \OVI\ column density versus \HI\ column density (\emph{Left}) and \HI\ b-value (\emph{Right}). Plotted for both WHIM (\emph{blue}) and warm (\emph{purple}) absorbers. Error bars show 1st and 4th quartiles. Dashed line indicates minimum b-value for a WHIM absorber, as given by Equation \eqref{eq:bth}.}
  \label{fig:OVI_in_HI}
\end{center}
\end{figure*}

One advantage of using the contour method is that all information 
associated with a single grid cell is also associated with a given 
lixel and consequently associated with a given absorber. This 
allows us to identify \HI\ absorbers and then automatically 
identify the associated \OVI\ content by summing the \OVI\ number 
densities along the lixels in the absorber. Thus, after identifying 
\HI\ absorbers using the contour method we find the column density 
of \OVI\ associated with this \HI\ identified set of cells. We plot 
this associated \OVI\ column density as a function of the original 
\HI\ number density and b-value in Figure \ref{fig:OVI_in_HI}. 

We find that the associated \OVI\ column shows very different 
behavior for WHIM absorbers ($T>10^5~K$) versus warm absorbers. In 
nearly all cases but the highest column \HI\ absorbers, the 
associated \OVI\ column density is significantly greater for WHIM 
absorbers than for warm absorbers. As a function of \HI\ column 
density the \OVI\ column for WHIM absorbers increases and then 
decreases, with the peak at around $10^{14}$~cm$^{-2}$. The 
median \OVI\ column for warm absorbers increases fairly steadily 
as a function of \HI\ column density. 

The median \OVI\ column density increases steadily as a function 
of b-value starting with the minimum possible b-value for WHIM 
absorbers. Warm absorbers show an initially sharp increase, but after 
$b\sim 60$~km~s$^{-1}$ the points become statistically insignificant. 
%The peak \OVI\ column density occurs at the $b\sim 40$~km~s$^{-1}$, where 
%the number of warm absorbers sharply decreases as the fraction of 
%total absorbers is dominated by WHIM absorbers.

\subsubsection{Metallicity Ionization Fraction Product}

The product $(Z/Z_\odot)\times f_{OVI}$, where $(Z/Z_\odot)$ is the gas
metallicity in solar units and $f_{OVI}$ is the fraction of oxygen
in the \OVI~ionization state, is of key importance for constraining
the budget of baryons traced by \OVI,  as given by 

\begin{equation}
  \Omega_b^{\mathrm{OVI}}=
  \left[
  \frac{\mu_bH_o}{c\rho_{cr}(O/H)_\odot}\right] \int_{N_{min}}^{N_{max}}
  \left(\frac{\partial^2\mathcal{N}}{\partial z\partial N}\right)\frac{N}{Z_O(N)f_{\mathrm{OVI}}(N)}dN
  \label{eq:baryonOVI}
\end{equation}

where $\mu_b$ is the mean baryon mass per hydrogen, $\rho_{cr}$ is
the cosmic closure density, and $(O/H)_\odot$ is the solar oxygen abundance.

Previously the product $(Z/Z_\odot)\times f_{OVI}$ has been
assumed to be constant, with typical estimates of $Z/Z_\odot=0.1$ and
$f_{OVI}=0.2$, giving a product of $0.02$. \citet{teppergarcia2011} first
exploited the advantage of simulations in calculating $\Omega_{\mathrm{OVI}}$
taking into account the ionisation fraction and metallicity of each
individual absorber, and 
previous analysis of our simulations using the cut method 
continued this study to find that
this product varied in a power-law distribution
proportional to $(N_{OVI})^\gamma,\, \gamma={0.7}$. Using this variation with an
observational fit of $\partial\mathcal{N}/\partial z \propto N^{-\beta},\, \beta=2.0$ \citep{danforth2008}, one
can constrain the baryon budget as follows: 
\begin{equation}
  \Omega_b^{\mathrm{OVI}}\propto \int_{N_{min}}^{N_{max}}
  N^{1-\beta-\gamma}dN  \; .
\end{equation}

\begin{figure}
  \includegraphics[width=0.45\textwidth]{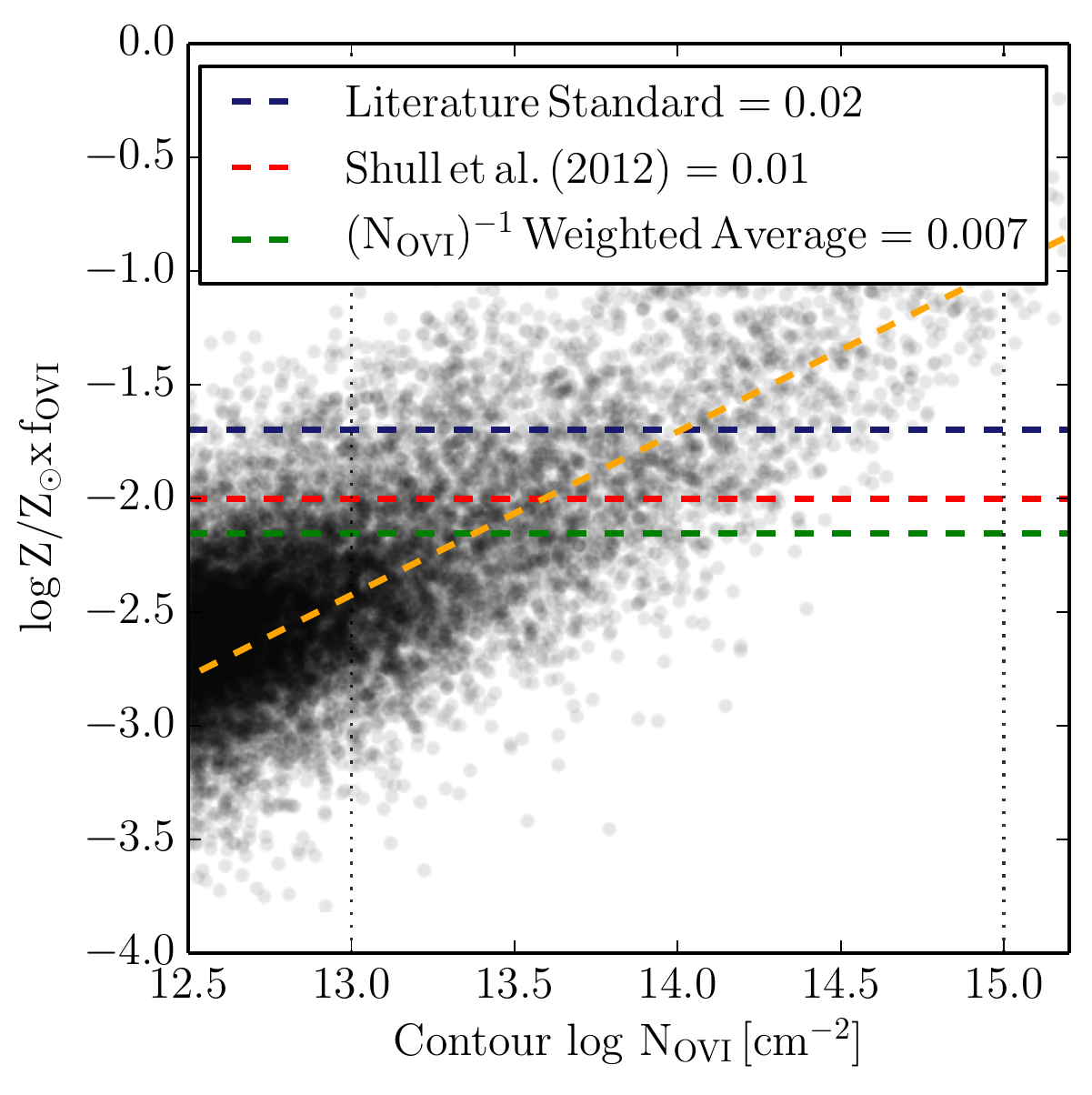}
  \caption{Values of the product $(Z/Z_\odot) \times f_{OVI}$ over column
    density.   The vertical gray dotted lines show limits of
    points that were included in averages and fit, representative of
    range of typical \OVI\ surveys \citep{danforth2008}. The
    horizontal blue dashed
    line shows the previously assumed value of 0.02
    obtained by using the standard values of $Z/Z_\odot=0.1$ and
    $f_{OVI}=0.2$. The horizontal red dashed line
    shows the average of column-limited points
    weighted by $(N_{OVI})^{-1}$. The horizontal green dashed line 
    shows value previously obtained in a similar $\log(N_{OVI})$
    weighted average by \citet{shull2012}.  The solid yellow line
    indicates a power law fit of the column-limited sample of
    points given
    by $(Z/Z_\odot)\times f_{OVI} = 0.021\times[N_{OVI}/(10^{14}
    \mathrm{cm}^{-2})]^{0.715}$. }
  \label{fig:zzf}
\end{figure}

As can be seen in Figure~\ref{fig:zzf}, we find a similar power-law 
distribution when using our more physically-motivated contour method
to find absorbers.  The product of ($Z/Z_{\odot}$) and $f_{OVI}$ is
proportional to $(N_{OVI})^{0.7}$. In an effort to make this result applicable
to observational surveys, we limit the absorbers we consider for the
average and fit to those with \OVI\ column densities between
$10^{13}$~cm$^{-2}$ and $10^{15}$~cm$^{-2}$. We then weight 
the product by $(N_{\mathrm{OVI}})^{1-\beta},\, \beta=2$ \citep{danforth2008} to estimate the
average contribution to the estimate 
$(Z/Z_\odot)f_{\mathrm{OVI}}=0.007$, which is once again a significantly lower value
than previous literature suggests.  As before, we only use absorbers within the
column density range $10^{13}$~cm$^{-2}\leq N_{\mathrm{OVI}}\leq 10^{15}$~cm$^{-2}$ to mimic
observational results. For further analysis of the implications of a lower value of the product $(Z/Z_\odot)\times f_{OVI}$ , we direct readers to \citet{shull2012}.

\section{Discussion}
\label{sec:discussion}

We have presented a reanalysis of a cosmological simulation designed
to study the intergalactic medium with the aims to both better understand the systematics of the methods used to identify absorbers and better understand the gas comprising the warm-hot intergalactic medium (WHIM; $10^5$~K$<T<10^7$~K) by analyzing the physical environments associated with the absorbers. We have devised three different methods for
identifying QSO absorption line systems in simulation data-- the cut
method, the contour method, and the spectral
method. The cut method slices the line of sight into regions of equal
redshift. The contour method identifies continuous regions above a
number density cutoff. The spectral method fits absorption spectra
generated for each line of sight. Often these methods give different
results, so it is important to choose the method with care.

\subsection{Absorber Identification Method Comparison}

Creating a one-to-one correspondence of absorbers between the two methods
allows us to get a sense of how well the two methods agree on column
density for an arbitrary absorber.  However, this method of matching is
imperfect and does not always match two absorbers that correspond to the same
physical structure. One way we combat this problem is by considering both
spectral absorber matching of individual line components as well as total absorbing
complexes. We do this to account for the cases where the spectral method
finds substructure that the contour method is unable to identify. 

Our spectral fitting routine does a good job of finding
absorbers in a manner that is consistent with the contour method for both
\lya\ and \OVI. Some fraction of \lya\ lines with column densities in the
neighborhood of $N_{HI}\sim 10^{16}$~cm$^{-2}$ are difficult
difficult to fit accurately due to line saturation. 

Results from the \dndz\ comparison for \lya\ absorbers show very good
agreement outside the $N_{HI}=10^{16}$~cm$^{-2}$ range between the
three methods. \OVI\ absorbers show considerable 
difference with observations, however, the \dndz\ values produced 
by the contour and spectral methods are quite consistent with each other. We hypothesize
that differences between simulated and observed \OVI\ absorption line statistics 
are due to ionization and feedback choices made in the underlying simulation, 
rather than in the methods we use to create and determine the processes of 
synthetic absorbers.

As the contour method does not take line of sight velocities into account
when identifying absorbers
we do not expect the b-values found by the spectral method to
instrinsically correspond to the b-values from the contour method. Despite
this difference, the two methods find comparable b-value distributions
to both each other and observations \citep{tripp2008} for \lya, especially
at lower column densities ($N_{HI}\leq 10^{15}$~cm$^{-2}$).  

There is a larger discrepancy between the contour method and the 
spectral method in \OVI\ which may be an indication of line of sight 
velocities having a larger effect or a systematic underfitting
by the spectral method. The slight systematic increase in the contour
method b-values at high column densities is consistent with observations \citep{tripp2008}
and some simulations \citep{2012ApJ...753...17C}, but contrast others
that find a much flatter distribution \citep{teppergarcia2011}. The b-value distribution 
of \OVI\ is also much more sensitive to the method of feedback used in 
the simulation indicating, so it is not instructive to read too much into
such a comparison.

One interesting result that comes from our comparison of the two
methods is the linear size-column density relation for both \HI\ and
\OVI.  We find (as can be seen in Figure~\ref{fig:size_Var}) that in
our calculations absorber size is a weak function of column density,
with $L \propto N^{-0.12}$ for both \HI\ and \OVI.  This is in
disagreement with analytic predictions
\citep[e.g.,][]{2001ApJ...559..507S}, though does not appear to
disagree with observations.  We speculate that this may be due to
simulation resolution or choice of cooling model, but also note that
the analytic models make the assumption of hydrostatic equilibrium,
which is not necessarily accurate.

\subsection{Understanding the WHIM}

The physical conditions of the absorbers showed roughly the
same results as found by \citet{britton2011}. The same general trends held
for mean \HI\ number density, metallicity, and \OVI\ fraction as a
function of column density. 

We also find a similar phase distribution of absorbers as \cite{britton2011}
with $69\%$ in the WHIM phase, $30\%$ in the warm phase, and
$1\%$ in the condensed phase. We do not, however, see strong evidence for a 
distinct bimodality in temperature of absorbers as \cite{britton2011} found for the phase
distribution of the total gas; instead we see a smoothly varying distribution across the temperature range. This does not indicate that there cannot
be two distinct collisionally ionized and photoionized populations, only that they are not represented in these absorber statistics. Such a population could still exist in the very rarest, highest column absorbers. It is also possible that these populations exist as part of a multi-phase absorber that would in effect, hide such a bimodality from observations. Further analysis could be done to examine such a distribution of gas within a single absorbing feature.

In effort to distinguish the characteristics of absorbers found in the WHIM
phase versus the warm phase, we segment the absorber population by temperature
cuts, with a temperature greater than $10^5$~K indicating that the
absorber is in the WHIM phase. These absorbers have similar 
column density distributions as the rest of the population, but b-values that are
much higher on average. There are no \HI\ (\OVI) absorbers in the WHIM phase with
b-values less than 40~km~s$^{-1}$ (10~km~s$^{-1}$) by definition, but above those 
limits the fraction of total absorbers is strongly dominated by WHIM absorbers.
These high b-value, low column \HI\ absorbers correspond to the population 
of broad \lya\ (BLA) absorbers identified through COS \citep{2010ApJ...710..613D}
and STIS \citep{2004ApJS..153..165R,2006A&A...451..767R}.

There has been substantial recent debate about the origin of \OVI\ absorption line systems, with some work suggesting that it comes entirely from collisional ionization, and other work suggesting that it comes from a combination of collisional and photo ionization.  As part of this debate, there has been controversy about the temperature of the plasma where \OVI\ absorption lines predominantly occur. In this work, we investigate the nature of \OVI\ absorbers in the intergalactic medium, but take a slightly different approach: we break the absorber population into absorbers dominated by collisional ionization and those dominated by photo ionization, and compare their thermal phase-space properties and column density distributions.  We find that the two populations have similar ranges of column densities, but (perhaps unsurprisingly) have bimodal distributions in both temperature and b-value.

To investigate the correlation between \HI\ and \OVI\ and the phases of gas
that they trace, we find the associated \OVI\ column densities for regions
identified as \HI\ absorbers. This shows a marked distinction between the
column of associated \OVI\ by column density and b-value of \HI\ for the
two phases, WHIM and warm, defined as having temperatures greater than and less thatn $10^5$~K respectively. The WHIM phase has a much higher median column density of associated \OVI\ in all regimes except the highest column \HI\ absorbers. This is 
consistent with the idea that the WHIM is effectively traced by low column, high b-value \lya\ absorbers (BLAs), as well as by \OVI.

Finally, in consideration of the product $(Z/Z_\odot)f_{\mathrm{OVI}}$, we find a
much lower average value than the literature standard, comparable with that
found by \cite{shull2012}. This may significantly affect predictions of IGM metallicity that have been primarily based on \OVI\ absorption.

\section{Summary}
\label{sec:summary}
In this paper, we compare two primary methods for finding
IGM absorbers along lines of sight cast through a simulation box.  One
method (the ``spectral method'') uses synthetic absorption line
spectra, and is meant to directly correspond to observational attempts
to find structure through fitting Voigt profiles to variations in
flux.  The other primary method (the ``contour method'') relies on
defining absorbers by associating contiguous regions along the line of
sight based on a threshold number density of the species of interest.
After comparing these methods to each other, we compare to observational data of \HI\ and \OVI\ absorption
line systems.  The key results of this paper are as follows:

\begin{enumerate}

\item The two methods give comparable column densities for a given
absorber, although there is some difficulty in creating a one-to-one
correspondence of absorbers between the two methods. The primary issue
for comparing absorbers generated with the two methods appears to be how 
one decides whether an absorber is a single coherent structure or a complex.

\item The number of \HI\ and \OVI\  absorbers per column density per unit redshift, or \dndz, traced by the two
methods give similar results, indicating that the two methods find
similar amounts of overall baryons in these ionization states regardless of the ability to match each
individual absorber. \HI\ \dndz\ compares favorably to observation,
while the \OVI\ \dndz\ in our simulation underpredicts the observational results.
This is likely a shortcoming of the simulation itself, as the abundance 
of OVI depends sensitively on the assumptions of star formation, feedback, and metal transport.

\item The distribution of Doppler parameters (or b-values) by column
density in our simulations are similar using the contour and spectral
methods in both \lya\ and \OVI.
Both methods of extracting b-values from our simulation match
observations of \lya\ systems as a function of column density. Neither method
provides a particularly good fit for the Doppler parameters measured in observed \OVI\ systems; we speculate that this 
is due to our choice stellar feedback algorithms.

\item The distribution of \OVI\ absorbers over baryon overdensity and
temperature was found to be similar to a previous analysis of the same simulation data,
but we do not see evidence for a bimodality in absorber distribution by temperature.
Instead, we see that \OVI\ absorbers are distributed smoothly in temperature-space 
from $4.5 < \log (T / K) < 5.5$.

\item Using the contour method, we find that the relationship between linear size and column density for a given absorber scales as L~$\propto$~N$^{-0.12}$, rather than the analytically-predicted exponent $-1/3$.  The reason for this is unclear, though we speculate that it may be related to simulation resolution or physics.

\item We examine the individual properties of warm vs. WHIM tracing \HI\ and \OVI\ absorbers (warm vs. WHIM having temperatures greater and less than $10^5$~K respectively) as well as photo- vs. collisional ionization dominated OVI absorbers.  \HI\ warm/WHIM-tracing absorbers show slightly different column density distributions with number of WHIM absorbers of a given column density falling off much more sharply than number of warm absorbers after $\mathrm{N_{HI}}\sim10^{14}$~cm$^{-2}$. \OVI\ show similarly shaped column density distributions, albeit with different normalizations.  \HI\ and \OVI\ absorbers associated with WHIM gas have systematically higher b-values.   \HI\ and \OVI\ absorbers associated with warm gas have b-value distributions centered around 12 and 10~km~s$^{-1}$, respectively, while \HI\ and \OVI\ absorbers associated with the WHIM have b-value distributions centered around 45 and 20~km~s$^{-1}$.  Dividing \OVI\ absorbers into photoionized  and collisionally ionized populations shows a similar results to the warm/WHIM division.

\item We investigate the association of warm and WHIM-tracing \HI\ absorbers with \OVI\ absorbers.  For \HI\ absorbers emanating from warm gas, there is a positive correlation between \HI\ column density and column density of associated \OVI.  Higher column density warm \HI\ absorbers tend to be associated with higher column density \OVI\ absorbers.  Comparatively, low column density WHIM-tracing \HI\ absorbers are associated with higher column density \OVI\ absorbers than are the warm-tracing \HI.  However, there does not exist such a trend of increasing \OVI\ column density with increasing \HI\ column density.  Instead, the average associated \OVI\ column density peaks at $\sim10^{13}$~cm$^{-2}$ at an \HI\ column density of $\sim 10^{14}$~cm$^{-2}$.

\item Finally, we study the relation between the column density of \OVI\ absorbers and the value of  $(Z/Z_\odot)\times
f_{\mathrm{OVI}}$, finding that $f_{\mathrm{OVI}}\,\propto\,
[N_{\mathrm{OVI}}/(10^{14}\,\mathrm{cm}^{-2})]^{0.7}$.  Over the column density range $13\leq \log N_{OVI}/ \mathrm{cm}^{-2} \leq 15$ this yields an average value of $(Z/Z_\odot)\times
f_{\mathrm{OVI}} = 0.007$, in reasonable agreement with \citet{shull2012}, but nearly a factor of three lower than earlier estimates.  This would imply that \OVI\ traces roughly triple the number of baryons previously thought.

\end{enumerate}

In subsequent papers we hope to expand our analysis of the systematic
differences in the contour and spectral methods, and use the nearly
one-to-one correlation between the two sets of absorbers to better
understand the correspondence between properties of absorption line
systems to features in physical structure. We also intend to look at
the effects of noise, as well as line blanketing when fitting multiple
ions together.

\acknowledgments

This work was supported by NASA through grants NNX09AD80G,
NNX12AC98G, NNX08AC14G, and NNX07AG77G,  and by the NSF through AST grant 0908819.  The simulations
presented here were performed and analyzed on the NICS Kraken and
Nautilus supercomputing resources under XSEDE allocations TG-AST090040
and TG-AST120009.  We thank Charles Danforth for helpful discussions
during the course of preparing this manuscript.  BWO was supported in
part by the MSU Institute for Cyber-Enabled Research and by Hubble Theory Grant  HST-AR-13261.01-A.  HE was
supported in part by the MSU Hantel Fellowship and the MSU College of
Natural Sciences.  \texttt{Enzo} and \texttt{yt} are developed by a
large number of independent research from numerous institutions around
the world.  Their commitment to open science has helped make this
work possible.

%\newpage

%\input{apj-bib}
\bibliographystyle{apj}
\bibliography{ms}

\begin{thebibliography}{}
\expandafter\ifx\csname natexlab\endcsname\relax\def\natexlab#1{#1}\fi

\bibitem[{Armstrong(1967)}]{armstrong1967}
Armstrong, B. 1967, Journal of Quantitative Spectroscopy and Radiative
  Transfer, 7, 61

\bibitem[{{Bristow} \& {Phillipps}(1994)}]{1994MNRAS.267...13B}
{Bristow}, P.~D., \& {Phillipps}, S. 1994, \mnras, 267, 13

\bibitem[{{Cen}(2012)}]{2012ApJ...753...17C}
{Cen}, R. 2012, \apj, 753, 17

\bibitem[{{Cen} \& {Chisari}(2011)}]{2011ApJ...731...11C}
{Cen}, R., \& {Chisari}, N.~E. 2011, \apj, 731, 11

\bibitem[{{Cen} \& {Fang}(2006{\natexlab{a}})}]{2006ApJ...650..573C}
{Cen}, R., \& {Fang}, T. 2006{\natexlab{a}}, \apj, 650, 573

\bibitem[{{Cen} \& {Fang}(2006{\natexlab{b}})}]{cen2006}
---. 2006{\natexlab{b}}, \apj, 650, 573

\bibitem[{{Cen} \& {Ostriker}(1992)}]{1992ApJ...399L.113C}
{Cen}, R., \& {Ostriker}, J.~P. 1992, \apjl, 399, L113

\bibitem[{{Cen} \& {Ostriker}(1999)}]{1999ApJ...514....1C}
---. 1999, \apj, 514, 1

\bibitem[{{Danforth} \& {Shull}(2005)}]{2005ApJ...624..555D}
{Danforth}, C.~W., \& {Shull}, J.~M. 2005, \apj, 624, 555

\bibitem[{{Danforth} \& {Shull}(2008)}]{danforth2008}
---. 2008, \apj, 679, 194

\bibitem[{{Danforth} {et~al.}(2006){Danforth}, {Shull}, {Rosenberg}, \&
  {Stocke}}]{2006ApJ...640..716D}
{Danforth}, C.~W., {Shull}, J.~M., {Rosenberg}, J.~L., \& {Stocke}, J.~T. 2006,
  \apj, 640, 716

\bibitem[{{Danforth} {et~al.}(2010){Danforth}, {Stocke}, \&
  {Shull}}]{2010ApJ...710..613D}
{Danforth}, C.~W., {Stocke}, J.~T., \& {Shull}, J.~M. 2010, \apj, 710, 613

\bibitem[{{Danforth} {et~al.}(2014){Danforth}, {Tilton}, {Shull}, {Keeney},
  {Stevans}, {Pieri}, {Stocke}, {Savage}, {France}, {Syphers}, {Smith},
  {Green}, {Froning}, {Penton}, \& {Osterman}}]{2014arXiv1402.2655D}
{Danforth}, C.~W., {Tilton}, E.~M., {Shull}, J.~M., {et~al.} 2014, ArXiv
  e-prints, arXiv:1402.2655

\bibitem[{{Dav{\'e}} {et~al.}(1999){Dav{\'e}}, {Hernquist}, {Katz}, \&
  {Weinberg}}]{1999ApJ...511..521D}
{Dav{\'e}}, R., {Hernquist}, L., {Katz}, N., \& {Weinberg}, D.~H. 1999, \apj,
  511, 521

\bibitem[{{Dav{\'e}} {et~al.}(2001){Dav{\'e}}, {Cen}, {Ostriker}, {Bryan},
  {Hernquist}, {Katz}, {Weinberg}, {Norman}, \& {O'Shea}}]{2001ApJ...552..473D}
{Dav{\'e}}, R., {Cen}, R., {Ostriker}, J.~P., {et~al.} 2001, \apj, 552, 473

\bibitem[{{Fang} \& {Bryan}(2001)}]{fang2001}
{Fang}, T., \& {Bryan}, G.~L. 2001, \apjl, 561, L31

\bibitem[{{Fukugita} {et~al.}(1998){Fukugita}, {Hogan}, \&
  {Peebles}}]{1998ApJ...503..518F}
{Fukugita}, M., {Hogan}, C.~J., \& {Peebles}, P.~J.~E. 1998, \apj, 503, 518

\bibitem[{{Fukugita} \& {Peebles}(2004)}]{2004ApJ...616..643F}
{Fukugita}, M., \& {Peebles}, P.~J.~E. 2004, \apj, 616, 643

\bibitem[{{Haardt} \& {Madau}(2001)}]{2001cghr.confE..64H}
{Haardt}, F., \& {Madau}, P. 2001, in Clusters of Galaxies and the High
  Redshift Universe Observed in X-rays, ed. D.~M. {Neumann} \& J.~T.~V. {Tran}

\bibitem[{Hopkins \& Beacom(2006)}]{0004-637X-651-1-142}
Hopkins, A.~M., \& Beacom, J.~F. 2006, The Astrophysical Journal, 651, 142

\bibitem[{{Norman} {et~al.}(2007){Norman}, {Bryan}, {Harkness}, {Bordner},
  {Reynolds}, {O'Shea}, \& {Wagner}}]{2007arXiv0705.1556N}
{Norman}, M.~L., {Bryan}, G.~L., {Harkness}, R., {et~al.} 2007, ArXiv e-prints,
  arXiv:0705.1556

\bibitem[{{Oppenheimer} \& {Dav{\'e}}(2009)}]{2009MNRAS.395.1875O}
{Oppenheimer}, B.~D., \& {Dav{\'e}}, R. 2009, \mnras, 395, 1875

\bibitem[{{Oppenheimer} {et~al.}(2009){Oppenheimer}, {Dav{\'e}}, \&
  {Finlator}}]{oppenheimer2009}
{Oppenheimer}, B.~D., {Dav{\'e}}, R., \& {Finlator}, K. 2009, \mnras, 396, 729

\bibitem[{{Oppenheimer} {et~al.}(2012){Oppenheimer}, {Dav{\'e}}, {Katz},
  {Kollmeier}, \& {Weinberg}}]{2012MNRAS.420..829O}
{Oppenheimer}, B.~D., {Dav{\'e}}, R., {Katz}, N., {Kollmeier}, J.~A., \&
  {Weinberg}, D.~H. 2012, \mnras, 420, 829

\bibitem[{{O'Shea} {et~al.}(2004){O'Shea}, {Bryan}, {Bordner}, {Norman},
  {Abel}, {Harkness}, \& {Kritsuk}}]{2004astro.ph..3044O}
{O'Shea}, B.~W., {Bryan}, G., {Bordner}, J., {et~al.} 2004, ArXiv Astrophysics
  e-prints, astro-ph/0403044

\bibitem[{{Persic} \& {Salucci}(1992)}]{1992MNRAS.258P..14P}
{Persic}, M., \& {Salucci}, P. 1992, \mnras, 258, 14P

\bibitem[{{Prochaska} \& {Tumlinson}(2009)}]{2009and..book..419P}
{Prochaska}, J.~X., \& {Tumlinson}, J. 2009, {Baryons: What,When and Where?},
  ed. H.~A. {Thronson}, M.~{Stiavelli}, \& A.~{Tielens}, 419

\bibitem[{{Rauch} {et~al.}(1997){Rauch}, {Miralda-Escude}, {Sargent}, {Barlow},
  {Weinberg}, {Hernquist}, {Katz}, {Cen}, \& {Ostriker}}]{1997ApJ...489....7R}
{Rauch}, M., {Miralda-Escude}, J., {Sargent}, W.~L.~W., {et~al.} 1997, \apj,
  489, 7

\bibitem[{{Richter} {et~al.}(2006){Richter}, {Fang}, \&
  {Bryan}}]{2006A&A...451..767R}
{Richter}, P., {Fang}, T., \& {Bryan}, G.~L. 2006, \aap, 451, 767

\bibitem[{{Richter} {et~al.}(2004){Richter}, {Savage}, {Tripp}, \&
  {Sembach}}]{2004ApJS..153..165R}
{Richter}, P., {Savage}, B.~D., {Tripp}, T.~M., \& {Sembach}, K.~R. 2004,
  \apjs, 153, 165

\bibitem[{{Savage} {et~al.}(2014){Savage}, {Kim}, {Wakker}, {Keeney}, {Shull},
  {Stocke}, \& {Green}}]{2014ApJS..212....8S}
{Savage}, B.~D., {Kim}, T.-S., {Wakker}, B.~P., {et~al.} 2014, \apjs, 212, 8

\bibitem[{{Schaye}(2001)}]{2001ApJ...559..507S}
{Schaye}, J. 2001, \apj, 559, 507

\bibitem[{Shull {et~al.}(2012)Shull, Smith, \& Danforth}]{shull2012}
Shull, J.~M., Smith, B.~D., \& Danforth, C.~W. 2012, \apj, 759, 23

\bibitem[{{Smith} {et~al.}(2008){Smith}, {Sigurdsson}, \&
  {Abel}}]{2008MNRAS.385.1443S}
{Smith}, B., {Sigurdsson}, S., \& {Abel}, T. 2008, \mnras, 385, 1443

\bibitem[{{Smith} {et~al.}(2011){Smith}, {Hallman}, {Shull}, \&
  {O'Shea}}]{britton2011}
{Smith}, B.~D., {Hallman}, E.~J., {Shull}, J.~M., \& {O'Shea}, B.~W. 2011,
  \apj, 731, 6

\bibitem[{{Tepper-Garc{\'{\i}}a} {et~al.}(2011){Tepper-Garc{\'{\i}}a},
  {Richter}, {Schaye}, {Booth}, {Dalla Vecchia}, {Theuns}, \&
  {Wiersma}}]{teppergarcia2011}
{Tepper-Garc{\'{\i}}a}, T., {Richter}, P., {Schaye}, J., {et~al.} 2011, \mnras,
  413, 190

\bibitem[{{The Enzo Collaboration} {et~al.}(2013){The Enzo Collaboration},
  {Bryan}, {Norman}, {O'Shea}, {Abel}, {Wise}, {Turk}, {Reynolds}, {Collins},
  {Wang}, {Skillman}, {Smith}, {Harkness}, {Bordner}, {Kim}, {Kuhlen}, {Xu},
  {Goldbaum}, {Hummels}, {Kritsuk}, {Tasker}, {Skory}, {Simpson}, {Hahn},
  {Oishi}, {So}, {Zhao}, {Cen}, \& {Li}}]{2013arXiv1307.2265T}
{The Enzo Collaboration}, {Bryan}, G.~L., {Norman}, M.~L., {et~al.} 2013, ArXiv
  e-prints, arXiv:1307.2265

\bibitem[{{Thom} \& {Chen}(2008{\natexlab{a}})}]{2008ApJS..179...37T}
{Thom}, C., \& {Chen}, H. 2008{\natexlab{a}}, \apjs, 179, 37

\bibitem[{{Thom} \& {Chen}(2008{\natexlab{b}})}]{2008ApJ...683...22T}
---. 2008{\natexlab{b}}, \apj, 683, 22

\bibitem[{{Tilton} {et~al.}(2012){Tilton}, {Danforth}, {Shull}, \&
  {Ross}}]{tilton2012}
{Tilton}, E.~M., {Danforth}, C.~W., {Shull}, J.~M., \& {Ross}, T.~L. 2012,
  \apj, 759, 112

\bibitem[{{Tripp} {et~al.}(2008){Tripp}, {Sembach}, {Bowen}, {Savage},
  {Jenkins}, {Lehner}, \& {Richter}}]{tripp2008}
{Tripp}, T.~M., {Sembach}, K.~R., {Bowen}, D.~V., {et~al.} 2008, \apjs, 177, 39

\bibitem[{{Turk} {et~al.}(2011){Turk}, {Smith}, {Oishi}, {Skory}, {Skillman},
  {Abel}, \& {Norman}}]{yt}
{Turk}, M.~J., {Smith}, B.~D., {Oishi}, J.~S., {et~al.} 2011, \apjs, 192, 9

\end{thebibliography}

\end{document}